\documentstyle[12pt,epsf,amstex]{article}
\advance\voffset by -5mm   
\advance\hoffset by -8mm  
\makeatletter
\@addtoreset{equation}{section}
\makeatother

\renewcommand{\title}[1]{\null\vspace{25mm}

\noindent{\Large{\bf #1}}\vspace{10mm}

\noindent {\large By }}
\newcommand{\authors}[1]{\noindent{\large #1}\vspace{3mm}

}
\newcommand{\address}[1]{\noindent #1\vspace{5mm}

}
\renewcommand{\abstract}[1]{\vspace{19mm}

\noindent{\small{\em Abstract.} #1}\vspace{2mm}

} 

\newtheorem{lemma}{Lemma} 
\newtheorem{lemmaA}{Lemma} 
\newtheorem{theorem}{Theorem}
\newtheorem{theoremA}{Theorem}

\newcommand{\C}{{\Bbb C}}

\newcommand{\R}{{\Bbb R}}

\newcommand{\Z}{{\Bbb Z}}
\newcommand{\cC}{{\cal C}}

\newcommand{\ga}{\alpha}
\newcommand{\gb}{\beta}

\newcommand{\gC}{\Gamma}
\newcommand{\gd}{\delta}
\newcommand{\gD}{\Delta}
\newcommand{\gep}{\varepsilon}

\newcommand{\gl}{\lambda}

\newcommand{\gn}{\nabla}
\newcommand{\go}{\omega}

\newcommand{\gt}{\theta}

\newcommand{\be}{\begin{eqnarray}}
\newcommand{\ee}{\end{eqnarray}}
\newcommand{\bee}{\begin{eqnarray*}}

\newcommand{\eea}{\end{array} $$}
\newcommand{\eeaa}{\end{array} \ee}
\newcommand{\eee}{\end{eqnarray*}}

\newcommand{\I}{{\rm i}}
\newcommand{\Int}{\displaystyle \int}
\newcommand{\intoi}{\Int_0^\infty}
\newcommand{\intii}{\Int_{-\infty}^{+\infty}}
\newcommand{\intio}{\Int_{-\infty}^0}

\newcommand{\LL}{ {\textstyle L^2(\frac{dp}{2p},\R_+) } }

\newcommand{\OO}{{\cal O}}

\newcommand{\saut}{\hspace{10mm}}

\newcommand{\sch}{{\cal S}(\R)}

\newcommand{\ssp}{{\cal S}(\R_+)}

\newcommand{\xh}{x_{_{Th}}}

\newcommand{\xph}{x'_{_{Th}}}
\newcommand{\xpph}{x''_{_{Th}}}

\newcommand{\ys}{y_{_A}}

\newcommand{\abs}[1]{\mid \! #1 \! \mid}
\newcommand{\bea}[1]{$$ \begin{array}{#1}}
\newcommand{\beaa}[1]{\be \begin{array}{#1}}
\newcommand{\ch}[1]{{#1
\hspace{-1.65mm}^{\rule[-2.8mm]{0mm}{4.5mm}{\wedge}} }}

\newcommand{\fa}[1]{{\ind{#1}{\ga}}}
\newcommand{\fy}[1]{{\ind{#1}{y_o}}}

\newcommand{\ignore}[1]{}
\newcommand{\ind}[2]{{#1 \! \! _{#2}}}

\newcommand{\Indice}[3]{\indsupspec{\ind{#1}{#3}}{#2}}
\newcommand{\indsup}[2]{\Indice{#1}{#2}{\ \ }}
\newcommand{\indsupspec}[2]{{#1 \!\!\!^{#2}}}
\newcommand{\LLd}[1]{{ L^2(d #1,\R) }}

\newcommand{\LLup}[1]{{ L^1(d #1,\R_+) }}
\newcommand{\Lu}[1]{{ \parallel \! #1 \!\parallel _{_{L^1}} }}
 
\newcommand{\no}[1]{{ \parallel \! #1 \!\parallel }}

\newcommand{\sur}[2]{\frac{\textstyle #1}{\textstyle #2}}

\newcommand{\tf}[1]{{{#1 \! \!}^{^\sim}}}
\newcommand{\tfa}[1]{{{#1 \! \! \!}^{^\sim}}}
\newcommand{\tfch}[1]{{\tfa{\ch{#1}}}}
\newcommand{\Tfch}[1]{{\tfa{\ch{#1}}\,}}

\newcommand{\wh}[1]{{\widehat{#1}}}
\newcommand{\tfpa}{\ind{\tf{f}^{\gd}\!\!}{p_o,y_o}}
\newcommand{\pa}{\ind{f^{\gd}\!}{p_o,y_o}}
\newcommand{\chpa}{\ind{\ch{f} {}^{\gd}\!}{p_o,y_o}}

\newcommand{\packet}{\ind{\tf{f}^{\gd}\!\!}{p_o}\,\,}
\newcommand{\packets}{\ind{\tf{f}^{o}\!\!}{p_o}\,\,}
\newcommand{\wave}{{f^{\gd}_{p_o}}}
\newcommand{\waves}{{f^{o}_{p_o}}}

\newcommand{\wapas}{\ind{f^{o}\!}{p_o,y_o}}

\begin{document}

\hfill{Imperial/TP/97-98/1} \\ [-10mm]

\title{Some exact results on \\ [4mm]
the CGHS black-hole radiation\footnote{
Work done towards a Ph.D. at Lausanne University.}}
\authors{F. Vendrell\footnote{
Supported by the Soci\'et\'e Acad\'emique Vaudoise and by the Swiss National 
Science Foundation.}}
\address{\noindent
Institut de physique th\'eorique,
Universit\'e de Lausanne \\
CH-1012 Dorigny, Switzerland
\\ [-2mm] \\ and \\ [-2mm]\\
Blackett Laboratory,
Imperial College  \\  
London SW7 2BZ, UK} 

\abstract{
Theorems on the emission of massless scalar particles by the CGHS black hole 
are presented.  The convergence of the mean number of particles created 
spontaneously in an arbitrary state is studied and shown to be strongly 
dependent on the infrared behavior of this state. 
A bound for this quantity is given and its asymptotic forms close
to the horizon and far from the black hole are investigated.
The physics of a wave packet is analysed in some detail in the black-hole 
background.  It is also shown that for some states the mean number of created 
particles is {\it not} thermal close to the horizon.  These states have a 
long queue extending far from the black hole, or are unlocalised in 
configuration space.
}

\section{Introduction}

The quantum physics of black holes has been a field of extensive
research since Hawking discovered that, due to quantum mechanical effects,
black holes emit spontaneously particles with a thermal spectrum \cite{Haw}.
In order to understand better the physical outcomes of this discovery, as
for example the evaporation and entropy of black holes, two-dimensional 
black-hole models have been studied for simplicity in the literature.  
One of these is the CGHS black hole \cite{CGHS} which is again 
considered here.

The present paper is mainly concerned by the study, in the CGHS black-hole 
background, of the mean number $\bar{N}[f]$ of massless scalar particles 
created spontaneously in an {\it arbitrary} state $f$.  The investigation of 
this issue has first been done by Wanders \cite{Wan} for the Dirac massless 
field.  In that case, this quantity is reinterpreted as the probability $W[f]$
of detecting a fermion in the state $f$. Wanders showed that this probability
tends to the thermal probability $W^{Th}_\beta[f]$ when the state $f$ is
translated towards the event-horizon, where $\beta$ is the inverse temperature 
of the black hole.  He obtained furthermore a bound for the difference 
$\vert\,W[f]-W^{Th}_\beta[f]\,\vert$, which exhibits a strong dependence on 
the queue of $f$ extending far away from the horizon.

The massless scalar field will be studied here along similar lines.  In this 
case, however, infrared issues are of primordial importance, so I shall 
also concentrated on them.  Although the Wightman function is in general
not positive definite in two-dimensional spacetimes because of its bad IR 
behaviour \cite{MPS}, the massless scalar field may still be considered if the
set of states is reduced in an appropriate way \cite{FR}. In this framework, 
one of the relevant problem is then the influence of the infrared behavior of 
the state $f$ on the mean number $\bar{N}[f]$ and on the difference 
$\vert\,\bar{N}[f]-\bar{N}^{Th}_\beta[f]\,\vert$, where $\bar{N}^{Th}_\gb[f]$ 
is the average number of particles in the state $f$ for an outgoing thermal 
flux of radiation of temperature $\gb^{-1}$.  It is shown in this paper that 
these quantities may diverge or not depending on the IR properties of the 
considered state.  This imply in particular that {\it there exist states for 
which the mean number of created particles is not thermal close to the 
horizon}.  The IR properties of $f$ are related to the properties of its queue
in configuration space. As in the fermionic case, a bound is obtained for the 
difference $\vert\,\bar{N}[f]-\bar{N}^{Th}_\beta[f]\,\vert$, which depends 
strongly on the queue of $f$ extending far away from the black hole.

The queues of states in configuration space play thus a relatively 
important role for the black-hole physics. When one restricts oneself to 
positive momentum modes only, states cannot be well localised.
There is a theorem of Paley and Wiener (see appendix \ref{sub:apth}) 
which asserts that if the Fourier transform of a function of one variable 
vanishes for all negative values of its argument, then it does not 
decrease at infinity faster that an exponential function. One is thus led to 
study states whose wave function decreases at infinity in an algebraic way or 
which are unlocalised.  One may expect that the global properties of these 
states come into play when the physics of the black hole is analysed close to 
its event-horizon. This issue is considered in the present paper.

The physics of a wave packet in the CGHS black-hole 
background is also studied.  This wave packet depends on a parameter $\delta$,
and its Fourier transform becomes narrower in momentum 
space when $\delta$ vanishes. In this limit, the variance of the momentum 
operator vanishes in this state and the wave packet is completely 
delocalised. In consequence, it is not justified to make the approximation of 
the horizon to calculate the mean number $\bar{N}[f]$, as it is
usually done, and an exact calculation must be performed. This is done here
for the first time and leads to unexpected results. In particular, the mean 
number of particles created by the black hole in this delocalised state is 
equal to the {\it half} of the thermal average $\bar{N}^{Th}_\beta[f]$, and 
is invariant under a translation of the state $f$.

Section 2 is devoted to a review of the CGHS black hole and of the
quantum field formalism \cite{Ve1}.  In section 3, some conditions 
under which the mean number $\bar{N}[f]$ of created particles diverges are 
investigated and a bound for this quantity is given.  The asymptotic 
behaviors of the mean number $\bar{N}[f]$ close to the horizon and far from 
the black hole are given in section 4.  These last results are applied to 
the physics of a wave packet in the black-hole background in \mbox{section 5}.

\section{Quantum field theory in the black-hole background}

\subsection{The CGHS black hole}

The CGHS black hole \cite{CGHS} is a vacuum solution of the dilatonic gravity 
theory defined by the action
\be
S &=& \sur{1}{2\pi}\, \int d^2x\,\sqrt{-g}\
\left\{\,e^{-2\phi}\,\left[\,R+4\,(\gn \phi)^2+4\,\gl^2 \,\right] \,
-\sur{1}{2}\,(\gn f)^2\,\right\},
\label{CGHSaction}
\ee
where $g$ is the metric, $\phi$ the dilatonic field, $f$ a
classical massless matter field and $\gl^2$ the cosmological constant.
This black hole may be created from a shock wave of $f$-matter, whose
only non-vanishing energy-momentum tensor $T^f_{\mu\nu}(x)$ component is
\beaa{rcccl}
T_{++}^f(x) &=& \sur{1}{2}\,(\partial_+ f)^2 &=& M\,\gd(x^+),
\label{CGHSTEIf}
\eeaa
where $M>0$.  For simplicity, one assumes that $\gl=M=1$ without loss
of generality.
If the line element is Minkowskian for $x^+\leq0$, then from
the equations of motion one gets for $x^+\geq0$
\be
ds^2 &=& \sur{dx^+dx^-}{1+\,e^{x^-}\,(\,e^{-x^+}-1\,)}\cdot
\label{CGHSmetricx}
\ee
The $x$ coordinates are the incoming coordinates, the outgoing coordinates
\mbox{$(y^+,y^-)\in\R^2$} are defined by the transformation 
\be
\left\{
\begin{array}{rcl}
x^+(y^+)&=& y^+, \\ [2mm]
x^-(y^-)&=& -\log \left(\,1+e^{-y^-}\,\right).
\end{array}\right.
\label{x(y)CGHS}
\ee
These coordinates parametrise only the lower half-plane $x^-<0$ of spacetime 
where the line element (\ref{CGHSmetricx}) is given by
\be
ds^2 &=& \sur{dy^+dy^-}{1+e^{y^--y^+}},
\ee
if $x^+\geq0$.
This tends to the Minkowski metric in the limit $y^+\rightarrow+\infty$.
The spacetime diagram in shown in fig.~1.
\begin{figure} 
\centerline{
\epsfysize=3.0truein
\epsfbox{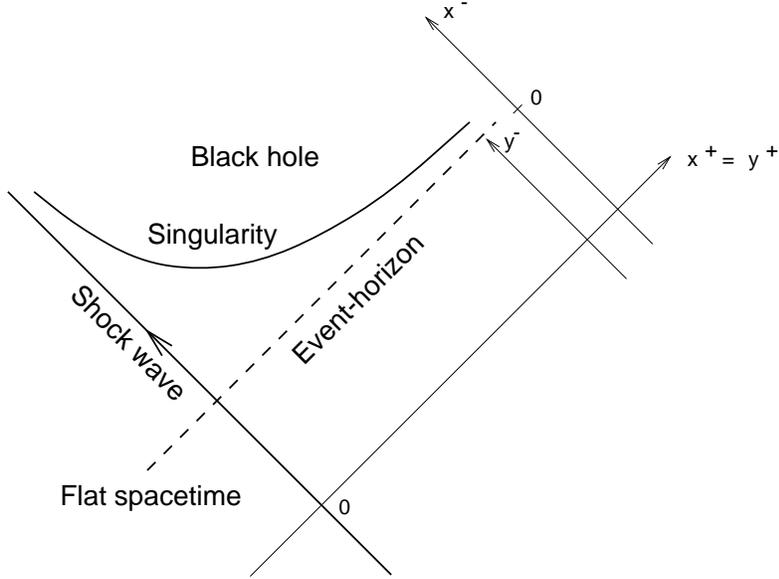}}
\vspace{5mm}
\caption{Spacetime diagram of the CGHS black hole in $x$ coordinates. 
The event-horizon is located at $x^-=0$ or $y^-=+\infty$.
}
\label{fig}
\mbox{}\vspace{0mm}
\end{figure}

Equation (\ref{CGHSmetricx}) implies that the scalar curvature is 
singular on the curve $x^-=x^-_S(x^+)$ where \cite{Ve3}
\be
x^-_S(x^+) &=& -\log\left(\,1-e^{-x^+}\,\right).
\ee
The signature of the line element (\ref{CGHSmetricx}) is reversed there, 
i.e.~the conformal factor is only positive if $x^-<x^-_S(x^+)$.
This line element may be rewritten in terms of the function $x^-_S$:
\be
ds^2 &=& \sur{dx^+\,dx^-}{1-e^{x^--x^-_S(x^+)}}\cdot
\label{eq:gui*}
\ee
In the Minkowskian region close to the singularity, i.e.~in the limits 
$x^-\approx x^-_S$ and $x^+\gg 1$, eq.~(\ref{eq:gui*}) tends to \cite{Ve1,Ve2}
\be
ds^2 &=& -\frac{dx^+\,dx^-}{x^-},
\label{eq:gui}
\ee
since $x^-_S(x^+)\approx0$ if $x^+\gg 1$. 
For this new line element, the transformation $x=x(y)$ is redefined by
\be
\left\{
\begin{array}{rcl}
x^+(y^+)&=& y^+, \\ [2mm]
x^-(y^-)&=& -e^{-y^-},
\end{array}\right.
\label{x(y)gui}
\ee
and this line element is Minkowskian in these new $y$ coordinates.  In the 
limit $x^-\approx 0$, the two transformations (\ref{x(y)CGHS}) and 
(\ref{x(y)gui}) coincide. 

\subsection{Quantum field theory}

\subsubsection{Fields and test functions}

In a two-dimensional spacetime, the line element can always be written in
a conformal form in an appropriate set of coordinates, at least locally.
In these coordinates, the left and right moving modes of the massless scalar 
field decouple. If the transformation relating the incoming and outgoing
coordinates, denoted by $x$ and $y$ respectively, takes the form 
$x^\pm=x^\pm(y^\pm)$, these modes do not mix up when the change of coordinates 
is made, i.e.~left (right) moving modes in incoming coordinates are still
left (right) moving modes in outgoing coordinates.  The physics of the left
moving modes is then trivial in the CGHS black-hole background (see eqs 
(\ref{x(y)CGHS})), and so we will concentrate from now on only on the right 
moving modes, and the subscripts $\pm$ will be dropped. 

The incoming and outgoing field distributions, denoted by $\phi$ and 
$\ch{\phi}$, are related through the equation $\phi[\ch{f}]=\ch{\phi}[f]$
\cite{Ve1}. The incoming test function $\ch{f}(x)$ is given in terms of the 
outgoing test function $f(y)$ by
\be
f(y) &=& x'(y)\,\ch{f}(x(y)), \qquad \forall\,y\in\R,
\label{eq:defChf}
\ee
or by $\tfch{f}=U\tf{f}$, where the kernel of the operator $U$ is defined by
\be
U(k,p) &=& \sur{1}{2\pi} \intii dy\ e^{-\I kx(y)}\ e^{\I py}.
\label{eq:defU}
\ee
The incoming and outgoing momenta are denoted by $k$ and $p$ respectively.
The outgoing wave function space is $\LL$, and is the completion of the 
set\footnote{$\theta$ denotes the step function defined by $\theta(p)=0$
if $p<0$ and $\theta(p)=1$ otherwise.}
\be
\ssp &=& \{\, \tf{f}\in\sch \, \mid\,\tf{f}(p)=\theta(p)\tf{f}(p),\,
\forall\,p\in\R\,\},
\ee
where $\sch$ is the Schwartz space.

In the CGHS black-hole background, one has from eqs (\ref{x(y)CGHS}) and 
(\ref{eq:defU}),
\be
U(k,p) &=& \sur{1}{2\pi}\, B(\I p-\I k+0^+,-\I p+0^+).
\label{eq:trounoirnoyauU}
\ee
Here $B$ is the beta function defined by \cite{GuW}
\be
B(q,r)\ = \ B(r,q)\ = \ \sur{\gC(q)\,\gC(r)}{\gC(q+r)} \
= \ \intii dy\,\sur{e^{qy}}{(1+e^y)^{q+r}},
\label{eq:sponatneTNbetadef}
\ee
where $\gC$ is the gamma function satisfying
\be
\left\vert\,\gC(\I p)\,\right\vert^2 &=& \sur{\pi}{p\,\sinh (\pi p)}\cdot
\label{eq:identitegamma}
\ee

\subsubsection{Local observables} \label{sec:localtrounoir}

The two-point function for the incoming vacuum is given in $y$ coordinates by 
\cite{Ve1}
\be
\widehat{W}\,(y,y')&=&
-\frac{1}{4\pi} \, \log\left[\,x(y')-x(y)+\I\,0^+\, \right].
\label{eq:localtnfonctionadeuxpoints}
\ee
In the CGHS black-hole background, it is periodic in the imaginary direction 
for all $y,y'\in\R$,
\be
\widehat{W}\,(y,y') &=& \widehat{W}\,(y,y'+\I\,2\pi n),
\saut \forall \, n \in \Z.
\ee
Since the thermal two-point function is given by \cite{Ve1}
\be
W^{Th}_\gb\, (y,y') &=&
- \sur{1}{4\pi} \log \left\{\, \sur{\gb}{\pi} \,
\sinh \left[\,\frac{\pi}{\gb}\,\left(y'-y+\I\,0^+\right)\,\right]\,\right\},
\label{2PFthermal} 
\ee
one obtains in this case
\be
\wh{W}\,(y,y') &\approx& W_{2\pi}^{Th}\, (y,y'),
\saut \mbox{{\it when} \ $y,y'\gg 1$,}
\label{eq:localtnKMS} \\ [2mm]
\wh{W}\,(y,y') &\approx& W_\infty^{Th}\, (y,y'),
\saut \mbox{{\it when} \ $-y,-y'\gg 1$,}
\label{eq:localtnKMSbis}
\ee
if two-point functions are considered as kernel of distributions on 
$\ssp\times\ssp$.
This means that, in outgoing coordinates, the incoming vacuum is a 
thermal state of temperature $(2\pi)^{-1}$ close to the horizon, and of 
temperature zero far from the horizon.
Since the energy-momentum tensor $\wh{T}(y)$ in $y$ coordinates may be 
obtained from the two-point function $\wh{W}\,(y,y')$, these results imply 
that $\wh{T}(y)$ is also thermal close the horizon and far from the 
CGHS black hole,
\be
\lim_{y\rightarrow +\infty}
\wh{T}(y) \ = \ T^{Th}_{2\pi}, \qquad
\lim_{y\rightarrow -\infty}
\wh{T}(y) \ = \ T^{Th}_{\infty},
\ee
where the thermal energy-momentum tensor of temperature $\beta^{-1}$ is given 
by
\be
T^{Th}_\beta &=& \sur{\pi}{12\beta^2}\cdot
\label{eq:thermalTEI}
\ee

\subsubsection{Mean number of created particles and implementability}

The mean number of particles created spontaneously in a normalised state
$\tf{f}\in\LL$ is given by \cite{Ve1}
\be
\bar{N}[f] &=& \intoi \frac{dk}{2k} \abs{\tfch{f}(-k)}^2.
\label{defN}
\ee
In the CGHS black-hole background one has from eq.~(\ref{eq:identitegamma}),
\be 
\left\vert\,U(-k,p)\,\right\vert^2
&=& \sur{1}{4\pi}\,\sur{k}{p\,(p+k)}\,
\sur{\sinh\pi k}{\sinh\pi p\,\sinh \pi(p+k)} \\ [2mm]
&\approx& \sur{1}{2\pi p}
\left[\,\sur{\gt(k)}{e^{2\pi p}-1}+\sur{\gt(-k)}{1-e^{-2\pi p}}\,\right],
\qquad \mbox{{\it if} \ $\vert k\vert\gg\vert p\vert+1$.}
\label{eq:trounoirUhorizonA}
\ee
The mean number of particles created in a given mode $f_p$, defined by
$\ind{\tf{f}}{p}(p')=2p'\delta(p-p')$ where $p>0$, is UV divergent in $k$,
\be
\bar{N}\,[\,f_p\,] &=& \infty.
\ee
The total mean number of created particles is clearly also infinite,
with an additional IR divergence in $p$.  The incoming and outgoing vacuums 
can thus not be related by an unitary transformation, i.e.~the problem is 
not implementable. 

\subsubsection{The thermal case}

In ref.~\cite{Ve1}, it is shown that the theory of a massless scalar field 
interacting with an outgoing thermal flux of radiation is equivalent to the 
theory of this field in the spacetime background defined by 
eq.~(\ref{eq:gui}), in the 
sense that expectation values of observables in the incoming vacuum are equal 
to their thermal averages in the outgoing Hilbert space. 
In particular, the incoming-vacuum two-point function and energy-momentum 
tensor are given everywhere in this spacetime by eqs (\ref{2PFthermal}) and 
(\ref{eq:thermalTEI}) with $\beta=2\pi$, i.e.~they coincide 
with the thermal averages.  Similarly, the mean number of particles 
spontaneously created in this spacetime and in the normalised state 
$\tf{f}\in\LL$ is given by $\bar{N}[f]=\bar{N}^{Th}_\beta[f]$ with 
$\beta=2\pi$, where $\bar{N}^{Th}_\beta[f]$ is the thermal average given by
\be
\bar{N}^{Th}_\beta[f] &=& \intoi \sur{dp}{2p} 
\frac{\abs{\tf{f}(p)}^2}{e^{\beta p}-1}\cdot
\label{Nthermal}
\ee
In the spacetime defined by eq.~(\ref{eq:gui}), the incoming test function 
will thus be denoted by $\Indice{\ch{f}}{\,Th}{\ \ }(x)$, which is defined by 
eq.~(\ref{eq:defChf}), where the transformation $x(y)$ is given by 
eq.~(\ref{x(y)gui}).
This spacetime is the dynamical counterpart of the 
$\eta-\xi$ spacetime of Gui \cite{gui}, which corresponds to the thermal 
{\it equilibrium} case.  
Since the CGHS line element (\ref{eq:gui*}) coincides with the line element 
(\ref{eq:gui}) in the region defined by $x^-\approx0$ and $x^+\gg 1$, the 
outgoing radiation emitted by the CGHS black hole is thermal in that region.  

\section{The mean number of created particles}

\subsection{Convergence of the mean number $\bar{N}[f]$ } 

A quick look at eq.~(\ref{defN}) shows that the mean number $\bar{N}[f]$ of 
particles created by the black hole may diverge or not depending on the 
infrared and ultraviolet behaviors of the incoming test function $\tfch{f}(k)$.
These behaviors are related to the properties of the outgoing test function 
$f(y)$ through eq.~(\ref{eq:defChf}). It is therefore interesting to study the
influence of $f(y)$ on the convergence of the mean number 
$\bar{N}[f]$ in the CGHS black-hole background, 
and in the spacetime defined by eq.~(\ref{eq:gui}) as
well, in order to understand the physics of the black hole close to the 
horizon.  Of particular interest are the infrared behavior of $\tf{f}(p)$ 
and the decreasing properties of $f(y)$ at infinity. Their influence on the 
function $\tfch{f}(k)$ is explored in the two following lemmas.

\begin{lemma} \label{lemma:f*}
Let $\tfa{\chi}\in\cC^\infty(\R)$ be a function satisfying
$\tfa{\chi}(0)\not=0$, vanishing at infinity and such that all its derivatives
vanish at the origin and at infinity. If one defines the function 
$f_\ga$ by $\fa{\tf{f}}(p) \equiv \gt(p)\,p^\ga\,\tfa{\chi}(p)$, 
$\forall\,p\in\R$, where $\ga>0$, then one has in the CGHS black-hole 
background
\be
\abs{\ind{\Tfch{f}}{\ga}(-k)} & \leq & 
C_\ga\,k^{q(\ga)}+\OO(k),
\hspace{40.5mm} \mbox{if \ $k \approx 0^+$,}
\label{eq:zeroTN} \\ [3mm]
\ind{\Tfch{f}}{\ga}(-k) &=&
\sur{C_{\ga,-}}{\sqrt{2\pi}}\sur{B(k)}{\ga\,(\log k)^\ga}
+\OO \left[\sur{1}{(\log k)^{1+\ga}}\right], 
\hspace{10mm} \mbox{if \ $k \gg 1$,} 
\label{eq:infiniTN}
\ee
and in the spacetime defined by eq.~(\ref{eq:gui})
\be
\Indice{\Tfch{f}}{\,Th}{\ga}(-k) &=&
\left\{ \rule{0mm}{14mm} \right. 
\nonumber \\ [-27mm]
&& \hspace{5mm}
\sur{C_{\ga,+}}{\sqrt{2\pi}}\sur{A(k)}{\ga\,(-\log k)^\ga}
+\OO \left[\sur{1}{(-\log k)^{1+\ga}}\right],
\hspace{4mm} \mbox{if \ $k\approx0^+$,}
\label{eq:zerohorizon} \\ [3mm]
&&\hspace{5mm} 
\sur{C_{\ga,-}}{\sqrt{2\pi}} \sur{B(k)}{\ga\,(\log k)^\ga}
+\OO \left[\sur{1}{(\log k)^{1+\ga}}\right],
\hspace{12mm} \mbox{if \ $k\gg1$,}
\label{eq:infinihorizon}
\ee
where the functions $A$ and $B$ are both bounded by above and below,
$C_\ga$ and $C_{\ga,\pm}$ are three constants depending on $\ga$, and 
where the function $q$ is defined by
\be
q(\ga) &=& \left\{
\begin{array}{ll}
\ga, \qquad & \mbox{if \ $0<\ga\leq1/2$,} \\ [2mm]
1/2, & \mbox{if \ $1/2<\ga$.}
\end{array} \right.
\label{defq}
\ee
\end{lemma}

\paragraph{Proof} See appendix \ref{sub:apf*}. \\ \\
This first lemma shows in a generic example that both the IR and UV behaviors
of $\tfch{f}(-k)$ are determined by the IR properties of $\tf{f}(p)$.  The
smoothness assumption on $\tfa{\chi}$ is necessary in order that the 
decreasing property of $f_\ga(y)$ at infinity is well determined for in that 
case one has \cite{thesis}
\be
f_\ga(y) &=& \left\{ \rule{0mm}{14mm} \right. 
\nonumber \\ [-27mm]
&& \hspace{5mm}
\sur{C_{\ga,-}}{y^{1+\ga}}+\OO\left(\frac{1}{y^{2+\ga}}\right),
\qquad \mbox{{\it if} \ $-y\gg1$,} 
\label{decreasef} \\ [2mm]
&& \hspace{5mm}
\sur{C_{\ga,+}}{y^{1+\ga}}+\OO\left(\frac{1}{y^{2+\ga}}\right),
\qquad \mbox{{\it if} \ $y\gg1$,} 
\label{decreasefbis}
\ee
where $C_{\ga,\pm}$ are the same constants as in lemma \ref{lemma:f*} (if
$\tfa{\chi}\not\in C^\infty(\R)$, the discontinuity of one of the derivatives
of $\fa{\tf{f}}(p)$ may imply that $f(y)$ decreases more slowly).
Equations (\ref{decreasef}) and (\ref{decreasefbis}) determine then the 
behavior of $\ind{\ch{f}}{\ga}(x)$ for $-x\gg1$ and $x\approx 0^-$ 
respectively through eq.~(\ref{eq:defChf}), on which the properties of 
$\ind{\Tfch{f}}{\ga}(-k)$ depend.  It can actually be shown, in a more general
context, that the IR behavior of $\tfch{f}(-k)$ depends solely on the 
behavior of $f(y)$ far from the black hole, and that the UV behavior of 
$\tfch{f}(-k)$ depends solely on the behavior of $f(y)$ close to the horizon. 
This is done in the following lemma which gives bounds instead of asymptotic 
behaviors for weaker assumptions, in particular it is assumed that the 
{\it modulus} of $f$ decreases at infinity faster than the inverse of an 
algebraic function.  From theorem \ref{th:causal}, if $f$ is squared 
integrable, it is always possible to choose the phase of $f$ in such a way 
that its Fourier transform $\tf{f}$ vanishes for all negative values of its 
argument.

\begin{lemma} \label{lemma:f**}
If $f$ is an integrable function satisfying $\tf{f}(0)=0$, whose derivative 
$f'$ exists and is integrable, and such that its modulus $\vert f\vert$ 
satisfies
\be
\abs{f(y)} &\leq&
\left\{ \rule{0mm}{13mm} \right. 
\nonumber \\ [-25mm]
&& \hspace{5mm} \sur{C_-}{(-y)^{1+\gep}}, 
\hspace{10mm} \mbox{if \ $y \leq -L$,} 
\label{eq:lemmaf**}\\ [3mm]
&& \hspace{5mm} \sur{C_+}{y^{1+\ga}},
\hspace{15mm} \mbox{if \ $y \geq L$,} 
\label{eq:lemmaf**bis}
\ee
where $C_\pm>0$, $L>1$, $\ga>0$ and $\gep>0$ are five constants, then
\be
\vert\,\tfch{f}(-k)\,\vert &\leq &
\left\{ \rule{0mm}{12mm} \right. 
\nonumber \\ [-21mm]
&& \hspace{5mm}
C_\gep\,k^{q(\gep)} + \OO(k),
\hspace{24mm} \mbox{if \ $k\approx 0^+$,} 
\label{lemmaf**A} \\ [3mm]
&& \hspace{5mm} \sur{2^{1+\ga} C_+}{\ga\,(\log k)^\ga} 
+ \OO\left(\sur{1}{\sqrt{k}}\right),
\hspace{11mm} \mbox{if \ $k\gg1$,}
\label{lemmaf**B}
\ee
and
\be
\vert\,\indsup{\Tfch{f}}{\!Th}(-k)\,\vert & \leq &
\left\{ \rule{0mm}{14mm} \right. 
\nonumber \\ [-25mm]
&& \hspace{5mm}
\sur{3\,C_-}{\gep\,(-\log k)^\gep} + \OO(k),
\hspace{16mm} \mbox{if \ $k\approx 0^+$,} 
\label{lemmaf**C} \\ [3mm]
&& \hspace{5mm} \sur{2^\ga C_+}{\ga\,(\log k)^\ga} 
+ \OO\left(\sur{1}{\sqrt{k}}\right),
\hspace{10mm} \mbox{if \ $k\gg1$,}
\label{lemmaf**D}
\ee
where the function $q$ is defined by eq.~(\ref{defq}), and where
$C_\gep>0$ is a constant which diverges when $\gep\rightarrow0^+$.
\end{lemma}

\paragraph{Proof} See appendix \ref{sub:apf**}.
\\ \\
Under the assumptions of this lemma, the IR and UV behaviors of 
$\indsup{\Tfch{f}}{Th}(-k)$ and the UV behavior of $\tfch{f}(-k)$ are thus at
least inversely proportional to a power of the logarithm of $k$, 
and $\tfch{f}(-k)$ decreases at least in an algebraic way in the IR region.
Lemmas \ref{lemma:f*} and \ref{lemma:f**} imply from eq.~(\ref{defN}) that the 
mean number $\bar{N}[f]$ of created particles is very sensitive to the IR 
properties of $\tf{f}(p)$ or to the asymptotic behavior of $f(y)$ at infinity.
This is highlight in the two following theorems.

\begin{theorem} \label{th:convN*}
Let $\tfa{\chi}\in\cC^\infty(\R)$ be a smooth function satisfying
$\tfa{\chi}(0)\not=0$, vanishing at infinity and such that all its derivatives
vanish at the origin and at infinity.
If $\fa{\tf{f}}(p) \equiv \gt(p)\,p^\ga\,\tfa{\chi}(p)$ (\,$\forall\,p\in\R$)
is a normalised wave function where $\ga>0$, then one has the equivalences
\be
\bar{N}[\,f_\ga\,] < \infty &\Longleftrightarrow& \ga > 1/2, \\ [3mm]
\bar{N}^{Th}_{2\pi}\,[\,f_\ga\,] < 
\infty &\Longleftrightarrow& \ga > 1/2, \\ [3mm]
\left\vert\,\bar{N}[\,f_\ga\,]-\bar{N}^{Th}_{2\pi}\,[\,f_\ga\,]\,\right\vert 
< \infty &\Longleftrightarrow& \ga > 1/2.
\label{eq:apcontreexempleAbis}
\ee
If $\ga\leq1/2$, $\bar{N}[\,f_\ga\,]$ is only UV divergent in the incoming 
momenta $k$, whereas $\bar{N}^{Th}_{2\pi}\,[\,f_\ga\,]$ is both IR and UV 
divergent in $k$, and 
$\left\vert\,\bar{N}[\,f_\ga\,]-\bar{N}^{Th}_{2\pi}\,[\,f_\ga\,]\,\right\vert$ 
is only IR divergent in $k$.
\end{theorem}

\paragraph{Proof}
This theorem follows from lemma \ref{lemma:f*} and eq.~(\ref{defN}).  
For example one has
\beaa{l}
\left\vert\,\ind{\Tfch{f}}{\ga}(-k)\,\right\vert^2
- \left\vert\,\Indice{\Tfch{f}}{\,Th}{\ga}(-k)\,\right\vert^2 
\ = \ \left\{
\begin{array}{ll} 
\sur{C_{\ga,-}^2\,A(k)^2}{\ga^2\,(\log k)^{2\ga}}
+\OO \left[\,\sur{1}{(\log k)^{1+2\ga}}\,\right], 
& \mbox{{\it if} \ $k\approx0^+$,}
\\ [6mm]
\OO \left[\,\sur{1}{(\log k)^{1+2\ga}}\,\right],
& \mbox{{\it if} \ $k\gg1$,}  
\end{array} \right.
\label{eq:trounoirdifferencehorizon}
\eeaa
which implies that the difference 
$\vert\,\bar{N}[\,f_\ga\,]-\bar{N}^{Th}_{2\pi}\,[\,f_\ga\,]\,\vert$ is IR 
convergent in $k$ if and only if $\ga>1/2$, and that it is UV convergent 
in $k$.
\hfill{$\Box$}

\begin{theorem}  \label{th:convN**}
Let $\tf{f}\in\LL$ be a normalised wave function such that $f$ and $f'$ exist 
and are integrable, and such that its modulus $\vert f\vert$ satisfies
\be
\abs{f(y)} &\leq& \left\{
\begin{array}{ll}
\sur{C}{(-y)^{1+\gep}}, \quad &\mbox{if \ $y \leq -L$,} \\ [4mm]
\sur{C}{y^{1+\ga}},&\mbox{if \ $y \geq L$,}
\end{array} \right.
\ee
where $C>0$, $L>1$, $\ga >1/2$ and $\gep>0$ are four constants.
Then
\be
\bar{N}[f] &<& \infty,
\ee
and if furthermore $\gep>1/2$,
\be
\bar{N}^{Th}_{2\pi}[f] &<& \infty.
\ee
\end{theorem}

\paragraph{Proof}
This theorem is proved in a straightforward way from lemma \ref{lemma:f**}.  
\hfill{$\Box$}
\\ \\
Theorem \ref{th:convN*} shows that the numbers $\bar{N}[f]$ and 
$\bar{N}^{Th}_{2\pi}[f]$, and their difference diverge if $\tf{f}(p)$ does not 
decrease sufficiently fast at the origin $p=0$.
Theorem \ref{th:convN**} shows that the convergence of $\bar{N}[f]$ depends
essentially on the asymptotic behavior of $f(y)$ {\it far} from the black 
hole, and that the convergence of $\bar{N}^{Th}_{2\pi}[f]$ depends on the 
behavior of $f(y)$ for both $y\gg1$ and $-y\gg1$.

\subsection{A bound for the mean number $\bar{N}[f]$ }

A bound for the difference
$\vert\,\bar{N}[f]-\sur{1}{2}\,\bar{N}^{Th}_{2\pi}[f]\,\vert$
is now given, as a first step towards obtaining a bound for the mean number
$\bar{N}[f]$.

\begin{lemma} \label{l:bound}
If $\tf{f}\in \LL$ is a normalised wave function such that $f$ exists
and is integrable, then
\be
\left\vert\,\bar{N}[f]-\sur{1}{2}\,\bar{N}^{Th}_{2\pi}[f]\,\right\vert &\leq&
\sur{1}{8\pi^2}\left\vert\,\Int_0^\infty dp'\,\tf{f}(p')^*\,t(p')^*
\Int_0^\infty dp\ P\,\sur{\tf{f}(p)\,t(p)}{p-p'}\,\right\vert  
\nonumber \\ [3mm] & & + \ 
C\,\left[\,\Int_0^\infty dp\,\sur{\abs{\tf{f}(p)}}{\sqrt{p\,(e^{2\pi p}-1)}}\, 
\left(\,1+p^2+\log\sur{1+p}{p}\right)\,\right]^2,
\label{eq:bound}
\ee
where $t$ is a complex function satisfying
\be
\vert\,t(p)\,\vert &=& \sqrt{\sur{2\pi}{p\,(e^{2\pi p}-1)}}\,,
\label{eq:t(p)}
\ee
and $C>0$ is a constant.
\end{lemma}

\paragraph{Proof} See appendix \ref{sub:aplbound}.
\\ \\
Although this bound is quite complicated, it contains useful informations 
which will be exploited in section 5.  The first term of this bound is the 
main UV contribution in momentum $k$ and stems from the values of the wave
function close to the horizon.  The $p^2$ contribution in the second term is 
the UV correction to the first term, this is needed because the wave function 
$f$ is not necessarily localised close to the horizon.  The logarithmic 
expression in the second term is the IR contribution in $k$ and stems from 
the values of the wave function far from the black hole.  The first 
contribution in the second term is due to the finite values of $k$. 

Some cruder but simpler bounds are given in the following theorem.

\begin{theorem} \label{th:bound}
If $\tf{f}\in \LL$ is a normalised wave function such that $f$ exists
and is integrable, then 
\be
\bar{N}[f] &\leq& C \Int_0^\infty \sur{dp}{2p} \, 
\sur{\abs{\tf{f}(p)}^2}{1-e^{-2\pi p}},
\label{eq:bound*} \\ [2mm]
\left\vert \, \bar{N}[f]- \bar{N}^{Th}_{2\pi}[f] \, \right\vert &\leq&
C\,\Int_0^\infty \sur{dp}{2p} \, 
\sur{\abs{\tf{f}(p)}^2}{1-e^{-2\pi p}},
\label{eq:bound**}
\ee
where $C>0$ is a constant.
\end{theorem} 

\paragraph{Proof}
This theorem follows from lemma \ref{l:bound}. The first term on the 
r.h.s.~of eq.~(\ref{eq:bound}) is bounded by applying a theorem on Hilbert 
transforms (see theorem \ref{th:hilbert}) which enables us to treat the 
principal value
\be
\left\vert\,\Int_0^\infty dp'\tf{f}(p')^*\,t(p')^*
\Int_0^\infty dp\ P\,\sur{\tf{f}(p)\,t(p)}{p-p'}\,\right\vert &\leq&
4\pi^2 \Int_0^\infty \sur{dp}{2p}\, 
\sur{\abs{\tf{f}(p)}^2}{e^{2\pi p}-1}\cdot
\label{eq:Bornetnbis}
\ee
Using the Cauchy-Schwartz inequality for the second term, one gets 
eqs (\ref{eq:bound*}) and (\ref{eq:bound**}) from eq.~(\ref{Nthermal}).
\hfill{$\Box$}
\\ \\
Theorems \ref{th:convN*} and \ref{th:bound} are in agreement, both
predict that the mean number $\bar{N}[\,f_\ga\,]$ and the difference
$\vert\,\bar{N}[\,f_\ga\,]-\bar{N}^{Th}_{2\pi}\,[\,f_\ga\,]\,\vert$ converge 
if $\ga>1/2$, where $f_\ga$ is defined in theorem \ref{th:convN*}.
From theorem \ref{th:bound}, it is clear that if $\bar{N}[f]$ and 
$\vert\,\bar{N}[f]-\bar{N}^{Th}_{2\pi}[f]\,\vert$ are infinite, 
they may only be IR divergent in $p$.  

\section{Asymptotic behaviors of the mean number $\bar{N}[f]$}

\subsection{Close to the horizon}
\label{sec:proximitehorizon}

The asymptotic behavior of the mean number $\bar{N}[f]$ close to the horizon 
is now investigated. The translation of the wave function $f$ by a quantity 
$y_o$ is first defined by
\be
f_{y_o} (y) &=& f(y-y_o), \qquad \forall\,y\in\R.
\label{eq:defgeneraledefyo}
\ee
Then one asks oneself the questions:
Does the mean number of created particles in the state
$f_{y_o}$ tend to its thermal average if $y_o \rightarrow +\infty$~? 
In another words, do we have
\be
\lim_{y_0 \rightarrow +\infty} 
\left(\,\bar{N}[\,f_{y_o}\,]-\bar{N}^{Th}_{2\pi}\,[\,f_{y_o}\,]\,\right)
&\stackrel{?}{=}&  0
\label{eq:apquestion}
\ee
And if the answer to this question is positive, how does 
$\bar{N}[\,f_{y_o}\,]$ tend to $\bar{N}^{Th}_{2\pi}\,[\,f_{y_o}\,]$ in this 
limit?
Notice that since a translation of $f$ implies only a global change of the 
phase of $\tf{f}$, the thermal average $\bar{N}^{Th}_{2\pi}\,[\,f_{y_o}\,]$ 
does not actually depend on $y_o$ (see eq.~(\ref{Nthermal})).

Theorem \ref{th:convN*} tells us that the answer to question 
(\ref{eq:apquestion}) may be negative, since from
eq.~(\ref{eq:apcontreexempleAbis}) there are wave functions $f$ such that
\be
\left\vert\,\bar{N}[\,f_{y_o}\,]-
\bar{N}^{Th}_{2\pi}\,[\,f_{y_o}\,]\,\right\vert 
&=& \infty,
\saut \mbox{$\forall\,y_o\in\R$,} 
\label{eq:apcontreexempleA}
\ee 
i.e., although the numbers $\bar{N}[\,f_{y_o}\,]$ and 
$\bar{N}^{Th}_{2\pi}\,[\,f_{y_o}\,]$ are both infinite in these cases, their 
differences are IR divergent. 
Thus, even if the wave function $f$ is translated towards the horizon, the 
mean number $\bar{N}[\,f_{y_o}\,]$ may not tend to the thermal average 
$\bar{N}^{Th}_{2\pi}\,[\,f_{y_o}\,]$.
One expects from theorem \ref{th:convN*} and eqs (\ref{decreasef}) and 
(\ref{decreasefbis}) that such wave functions should decrease more slowly than
$1/\vert y\vert^{3/2}$ at infinity.  It turns out, however, that only the 
asymptotic behavior of $f(y)$ far from the black hole (i.e. for 
$y\rightarrow-\infty$) determines whether the mean number 
$\bar{N}[\,f_{y_o}\,]$ does tend or not to the thermal average 
$\bar{N}^{Th}_{2\pi}\,[\,f_{y_o}\,]$ close to the horizon, and if it does, 
how it does it. It is shown below that if $f(y)$ decreases strictly faster 
than $1/\vert y\vert^{3/2}$ far from the black hole, then the answer to 
question (\ref{eq:apquestion}) is positive.
Two bounds for the Fourier transform $\fy{\tfch{f}} (-k)$ are first given
in the following lemmas.

\begin{lemma} \label{lemma:AN*}
If $f$ is an integrable function such that $\tf{f}(0)=0$ and if its
modulus $\vert f\vert$ satisfies
\be
\abs{f(y)} &\leq& \sur{C}{\abs{y}^{1+\ga}},
\hspace{10mm} \mbox{if \ $y\,\leq-L$,}
\label{eq:dalgebrique*} 
\ee 
where $C$, $L$ and $\ga$ are three positive constants, then
\be
\sqrt{2\pi} \abs{\fy{\tfch{f}} (-k)} &\leq& 
\sqrt{k} \, e^{-y_o/2} \, \Lu{f}
+\sur{4C}{\ga}\,\sur{2^\ga}{\left(\,y_o-\log k-1\,\right)^\ga},
\label{eq:inea} 
\ee
where $0\leq k\leq e^{y_o-2L}$.  This result is also true in the 
thermal case, i.e.~for $\Indice{\tfch{f}}{Th}{y_o}(-k)$.
\end{lemma}

\paragraph{Proof} See appendix \ref{ap:AN*}. 

\begin{lemma} \label{lemma:AN**}
If the function $\gd\ch{f}$ is defined by the difference 
$\gd\ch{f}(x)\equiv\ch{f}(x)-\indsup{\ch{f}}{\,Th}(x)$ ($x<0$) 
where $f$ and its derivative are integrable, then
\be
\sqrt{2\pi} \abs{\fy{\widetilde{\gd \ch{f}}} (-k)} &\leq& 
\left(\,\sur{1}{\sqrt{k}}+\sur{1}{k}\,\right) \, 
\left( \, \Lu{f}+\Lu{f'} \right),
\label{eq:ineb} 
\ee
where $k>0$.
\end{lemma}

\paragraph{Proof} See appendix \ref{ap:AN**}. 
\\ \\
A bound for the difference $\left\vert\,\bar{N}[\,f_{y_o}\,] 
-\bar{N}^{Th}_{2\pi}\,[\,f_{y_o}\,]\,\right\vert$
is now given in terms of $y_o>0$.

\begin{theorem} \label{th:approche}
If $\tf{f}\in\LL$ is a normalised wave function such that $f$ and its 
derivative exist and are integrable, and if the modulus $\vert f\vert$ 
satisfies
\be
\abs{f(y)} &\leq& \sur{C}{\abs{y}^{1+\ga}},
\hspace{10mm} \mbox{if \ $y\,\leq-L$,}
\label{eq:dalgebrique} 
\ee 
where $C>0$, $L\geq 1$ and $\ga>1/2$ are three constants, then
\beaa{l}
\left\vert\,\bar{N}[\,f_{y_o}\,] -
\bar{N}^{Th}_{2\pi}\,[\,f_{y_o}\,]\,\right\vert \leq
\sur{1}{\ga^2(2\ga-1)}\, \sur{32\,C^2}{(y_o/4+L-1)^{2\ga-1}}
+ e^{4L-y_o/4} \left(\,\Lu{f}+ \Lu{f'} \right)^2
\label{eq:approche}
\eeaa
where $y_o>0$.
\end{theorem} 

\paragraph{Proof}  
Since the Fourier transforms $\tfch{f}(-k)$ and $\indsup{\tfch{f}}{\,Th}(-k)$ 
behave similarly for $k\gg 1$ but differently for 
$k\approx 0^+$ (see lemmas \ref{lemma:f*} and \ref{lemma:f**}), one writes
\be
\left\vert\,\bar{N}\,[\,f_{y_o}\,]-
\bar{N}^{Th}_{2\pi}\,[\,f_{y_o}\,]\,\right\vert
\ \leq \hspace{110mm}\nonumber \\ [2mm]
\Int_0^W \sur{dk}{2k} \,
\left[\,\abs{\ind{\tfch{f}}{y_0}(-k)}^2 
+\abs{\Indice{\tfch{f}}{Th}{y_0}(-k)}^2\,\right] 
 + \Int_W^\infty \sur{dk}{2k} \,
\left\vert \, \abs{\ind{\tfch{f}}{y_0}(-k)}^2 
-\abs{\Indice{\tfch{f}}{Th}{y_0}(-k)}^2 \, \right\vert, \ \ \
\label{eq:apAresultat}
\ee
where $0<W\leq e^{y_o-2L}$. Notice that
the bounds of both the integrals and integrands depends on $y_o$.
Lemma \ref{lemma:AN*} implies that
\beaa{l}
\Int_0^W \sur{dk}{2k} \,
\left[\,\abs{\ind{\tfch{f}}{y_0}(-k)}^2
+ \abs{\Indice{\tfch{f}}{Th}{y_0}(-k)}^2 \, \right] \ \leq
\hspace{80mm}\\ [4mm]\hspace{50mm}
\sur{W}{2}\,e^{-y_o}\,\no{f}_{L^1}^2 + 
\sur{4^\ga}{\ga^2\,(2\ga-1)}\,
\sur{16\,C^2}{\left(\,y_o-\log W-1\,\right)^{2\ga-1}}\cdot
\label{eq:apBresultat}
\eeaa
If $\gd \fy{\ch{f}}(x)\equiv\fy{\ch{f}}(x)-\Indice{\ch{f}}{\,Th}{y_o}(x)$ 
one has
\be
\left\vert\,\abs{\ind{\tfch{f}}{y_0}(-k)}^2 
-\abs{\Indice{\tfch{f}}{Th}{y_0}(-k)}^2\,\right\vert
&\leq& 2 \ \Lu{f} 
\abs{\fy{\widetilde{\gd \ch{f}}} (-k)}
+\abs{\fy{\widetilde{\gd \ch{f}}} (-k)}^2.
\ee
Applying lemma \ref{lemma:AN**} one gets if $k>0$,
\be
\left\vert\,\abs{\ind{\tfch{f}}{y_0}(-k)}^2 
-\abs{\Indice{\tfch{f}}{Th}{y_0}(-k)}^2\,\right\vert
&\leq& \sur{1}{2}\,\left( \sur{1}{\sqrt{k}}+\sur{1}{k}+ \sur{1}{k^2}\,\right)\,
\left(\ \no{f}_{L^1}+\no{f'}_{L^1}\,\right)^2.
\ee
This last equation implies
\be
\Int_W^\infty \sur{dk}{2k} \,
\left\vert \, \abs{\ind{\tfch{f}}{y_0}(-k)}^2 
-\abs{\Indice{\tfch{f}}{Th}{y_0}(-k)}^2 \, \right\vert \,\leq \,
\sur{1}{4}\,\left( \sur{2}{\sqrt{W}}+\sur{1}{W}+ \sur{1}{2W^2}\,\right)
\,\left(\,\no{f}_{L^1}+\no{f'}_{L^1}\,\right)^2. \ \ \ \
\label{eq:apCresultat}
\ee
From eqs (\ref{eq:apBresultat}) and (\ref{eq:apCresultat}) one gets
eq.~(\ref{eq:approche}) if $W= e^{y_o/2-2L}$ 
(which satisfies $W<e^{y_o-2L}$).
\hfill{$\Box$} 
\\ \\
This last theorem shows clearly that the mean number $\bar{N}[\,f_{y_o}\,]$ 
does tend to the thermal average $\bar{N}^{Th}_{2\pi}\,[\,f_{y_o}\,]$ when 
$y_o\rightarrow+\infty$ if $f$ decreases sufficiently fast far from the 
black hole.

\subsection{Far from the horizon}

The asymptotic behavior of the mean number $\bar{N}[f]$ far from the 
horizon is now investigated.
In this region, i.e.~in the limit $y\rightarrow -\infty$, the metric in
$y$ coordinates tends to the Minkowski metric, and thus there is no local
creation of particles there.  However, this does not imply that
the mean number $\bar{N}[f]$ is arbitrary small if the test function $f$ is 
translated towards that region, because one expects that the queue of $f(y)$ 
close to the horizon may have a significant contribution to $\bar{N}[f]$ even 
in that limit.  Theorem \ref{th:convN*} tells us that there are indeed wave 
functions $f$ such that
\be
\bar{N}\,[\,f_{y_o}\,] &=& \infty, \qquad \forall\,y_o\in\R, 
\label{eq:apcontreexempleB}
\ee
and that these wave functions decrease more slowly than $1/\vert y\vert^{3/2}$ 
at infinity.
One ask thus oneself the questions:
If $f_{y_o}$ is defined by eq.~(\ref{eq:defgeneraledefyo}), under what
conditions does one have
\be
\lim_{y_0 \rightarrow -\infty} \bar{N}[\,f_{y_o}\,] \ 
&\stackrel{?}{=}&0,
\label{eq:apquestionflat}
\ee
and if the answer to this question is positive, how does 
$\bar{N}[\,f_{y_o}\,]$ vanish in this limit?  
It is shown below that if $f(y)$ decreases strictly faster than 
$1/y^{3/2}$ close to the horizon, then the answer to question 
(\ref{eq:apquestionflat}) is positive.
Three bounds for the Fourier transform $\fy{\tfch{f}} (-k)$ are first given
in the following lemma.

\begin{lemma} \label{l:flat*}
Let $f$ be an integrable function such that its modulus $\vert f\vert$ 
satisfies
\be
\abs{f(y)} &\leq& \sur{C}{y^{1+\ga}},
\saut \mbox{if \ $y \ \geq L$,}
\label{eq:dalgebriqueplat*} 
\ee  
where $C>0$, $L\geq1$ and $\ga>0$ are three constants, and assume that
$y_o<0$. 
\begin{enumerate}
\item[a)] If $\tf{f}(0)=0$ and $\ga>1/2$, then
\be
\sqrt{2\pi}\,\abs{\ind{\tfch{f}}{y_o}(-k)} 
&\leq& 22\,C\,\sqrt{k} +k\,e^L\,\no{f}_{L^1},
\label{eq:flat*A}
\ee
where $e^{-L}\geq k\geq 0$.
\item[b)] If $f$ is such that $\tf{f}(p)=\theta(p)\,\tf{f}(p)$, 
$\forall\,p\in\R$, then
\be
\sqrt{2\pi}\,\abs{\ind{\tfch{f}}{y_o}(-k)} &\leq&
k\,\no{f}_{L^1}\,e^{L+y_o/2}+\sur{2\,C}{\ga}\,\sur{1}{(L-y_o/2)^\ga},
\label{eq:flat*B}
\ee
where $k\geq0$.
\item[c)] If the derivative of $f$ is integrable, then
\be
\sqrt{2\pi}\,\abs{\ind{\tfch{f}}{y_o}(-k)} &\leq&
\sur{2^{1+\ga}\,C}{\ga\,(\log k-2y_o-2)^\ga}
+\sur{2}{\sqrt{k}}\,
\left(\,2^{1+\ga}\,C+\no{f}_{L^1}+\no{f'}_{L^1}\,\right) \hspace{10mm}
\label{eq:flat*C}
\ee
where $k\geq2\,e^{2L}$.
\end{enumerate}
\end{lemma}

\paragraph{Proof} See appendix \ref{ap:flat*}.
\\ \\
A bound for  $\bar{N}[\,f_{y_o}\,]$ is now given in terms of $y_o<0$.

\begin{theorem} \label{th:flat}
If $\tf{f}\in\LL$ is a normalised wave function such that $f$ exists and
is integrable, and if the modulus $\vert f\vert$ satisfies
\be
\abs{f(y)} &\leq& \sur{C}{y^{1+\ga}},
\saut \mbox{if \ $y \ \geq L $,}
\label{eq:dalgebriqueplat} 
\ee  
where $C>0$, $L\geq1$ and $\ga>1/2$ are three constants, then
\be
\bar{N}\,[\,f_{y_o}\,] &\leq&  
\sur{1}{\ga\,(2\ga-1)}\,\sur{8\,C^2}{(-y_0/4+L/2-1)^{2\ga-1}} 
\nonumber \\ [3mm] & & 
+\ 20\,e^{4L+y_o/2} 
\left(\,2^{1+\ga}\,C +\Lu{f}+ \Lu{f'} \right)^2,
\label{eq:plat}
\ee
where $y_0 \leq -4\,(1+L)$.
\end{theorem}

\paragraph{Proof}
Lemma \ref{l:flat*} is applied.
The bounds (\ref{eq:flat*A}), (\ref{eq:flat*B}) and (\ref{eq:flat*C}) are 
useful for small, finite and large values of $k$ respectively.
If $w$ and $W$ are two constants such that $0<w\leq e^{-L}$ and 
$2e^{2L}\leq W$, 
then
\be
\int_0^w \sur{dk}{2k} \abs{\ind{\tfch{f}}{y_o}(-k)}^2 &\leq& 
\sur{242\,C^2}{\pi}\,w+\sur{1}{4\pi}\,e^{2L}\,w^2\,\no{f}_{L^1}^2,
\\ [2mm]
\int_w^W \sur{dk}{2k} \abs{\ind{\tfch{f}}{y_o}(-k)}^2 &\leq&
\sur{1}{4\pi}\,\no{f}_{L^1}^2\,e^{2L+y_o}\,W^2 
+\sur{2\,C^2}{\pi}\,\sur{1}{\ga^2\,(L-y_o/2)^{2\ga}}\,\log\sur{W}{w},
\\ [2mm]
\int_W^\infty \sur{dk}{2k} \abs{\ind{\tfch{f}}{y_o}(-k)}^2
&\leq& \sur{2\,C^2}{\pi}\,\sur{4^\ga}{\ga^2\,(2\,\ga-1)}\,
\sur{1}{(\log W-2y_o-2)^{2\ga-1}} 
\nonumber\\[1mm]& &
+\sur{2}{\pi}\,\sur{1}{W}\,
\left(\,2^{1+\ga}\,C+\no{f}_{L^1}+\no{f'}_{L^1}\,\right)^2.
\ee
These bounds imply eq.~(\ref{eq:plat}) if $w=e^{y_o/4}$ and $W=e^{L-y_o/4}$
under the constraint $y_o\leq -4\,(1+L)$, so that the assumptions stated
on $w$ and $W$ are satisfied.
\hfill{$\Box$}
\\ \\
This last theorem shows clearly that $\bar{N}[\,f_{y_o}\,]$ does vanish when 
$y_o\rightarrow-\infty$ if $f$ decreases sufficiently fast close to the 
horizon.

\section{A wave packet}

The physics of a wave packet in the CGHS black-hole background is now
considered. This wave packet depends on two parameters $\delta$ and 
$p_o$ and is defined by
\be
\packet(p) &=& \gt(p)\,\sqrt{2\,\vert p\vert} \ \gD_\gd (p-p_o),
\qquad \forall\,p\in\R,
\label{eq:packet}
\ee
where $p_o\geq\gd>0$ and
\begin{enumerate}
\item[$i)$] $\gD_\gd$ is a normalised function in $\LLd{p}$: 
$\int_{-\infty}^{+\infty} dp \, \abs{\gD_\gd(p)}^2 \ = 1$;
\item[$ii)$] the support of $\gD_\gd$ is included in the interval 
$(-\gd, \gd)$;
\item[$iii)$] $\packet\in\cC^r(\R)$ where $r\geq0$; 
$\left(\packet\right)^{(n)}$ vanish at the origin and at infinity for 
$n=0,1,..,r$; the derivatives 
$\left(\packet\right)^{(r+1)}$ and $\left(\packet\right)^{(r+2)}$ 
exist almost everywhere and are bounded and integrable respectively;
\item[$iv)$] the function $\gD_\gd$ is real and positive;
\item[$v)$] the function $\wave$ and its derivative are integrable.
\end{enumerate}
From property $i)$ the wave function $\packet$ is normalised in $\LL$, and  
property $ii)$ implies that $\packet$ is centred about $p_o$ in momentum 
space.
From property $iii)$ one shows that the function $\wave$ decreases in 
configuration space at least as $1/\vert y\vert^{r+2}$ if $\vert y\vert\gg1$,
and property $iv)$ implies that $\wave$ is centred about $y=0$.
Property $v)$ will be useful below, when theorems \ref{th:approche}
and \ref{th:flat} are applied to this wave function.  
The function $\pa$ is defined to be the translation of $\wave$ as in 
eq.~(\ref{eq:defgeneraledefyo}), and its Fourier transform is given by
\be
\tfpa(p) &=& \gt(p)\,\sqrt{2\,\vert p\vert} \ \gD_\gd (p-p_o) \ e^{-\I py_o},
\qquad \forall\,p\in\R.
\label{eq:packet*}
\ee
For example, if $\gD_\gd$ is the triangle-shaped function
\be
\gD_\gd (p) &=& \sqrt{\sur{3}{2\,\gd^3}}\ \left[\,\gt(p)\,\gt(\gd-p)\,(\gd-p)
+ \gt(-p)\,\gt(\gd+p)\,(\gd+p)\,\right],
\ee
conditions $i)$ to $v)$ with $r=0$ are satisfied.  In this case, the wave 
function in configuration space is given by
\be
\pa (y) &=& 4\,\sqrt{\sur{3 p_o}{2 \pi\,\gd^3}} \
e^{\I p_o (y-y_o)} \, 
\sin^2 \left[\, \sur{\gd\,(y-y_o)}{2}\,\right] \sur{1}{(y-y_o)^2} \,  
+ \gd \, {\cal O} \left( \, \sqrt{\sur{\gd}{p_o}}\,\right),
\label{eq:pa}
\ee
and takes its maximum value at $y=y_o$ where 
\be
\pa(y_o) &=& \sqrt{\sur{3\,\gd\,p_o}{2 \pi}} +
\gd \, {\cal O} \left( \, \sqrt{\sur{\gd}{p_o}}\, \right).
\ee

The wave length of the generic wave packet (\ref{eq:packet}) is given 
approximately by 
\be
\gD y &=& \sur{2\pi}{p_o}\cdot
\ee
If $\gD x$ is the wave length of the incoming wave packet $\chpa(x)$, one has
\be
\gD y\, (x) &=& y\, (x+\gD x)- y\,(x)\,.
\ee
The incoming momentum is approximately given by $k_o = 2\pi/\gD x$.
If $k_o \gg 1$, the incoming and outgoing momenta are related 
by \cite{AM} 
\be
p_o &\approx& (1-e^x) \, k_o, \saut \mbox{{\it if} \ $x<0$}.
\ee
Close to the horizon, one has $p_o \approx (-x) \, k_o$ and thus the outgoing 
momenta $p_o$ is strongly shifted towards the IR region when $k_o$ is kept 
fixed.  In the limit $x \rightarrow -\infty$, the incoming and outgoing
momenta are asymptotically equal.  One also has if $p_o \gg 1$,
\be
k_o &\approx& (1+e^y) \, p_o,
\ee
and thus close to the horizon the incoming momenta $k_o$ is strongly shifted 
towards the UV region when $p_o$ is kept fixed.

The limits $\gd\rightarrow 0^+$ and $y_o\rightarrow\pm\infty$ of the mean
number of particles $\bar{N}\,[\,\pa\,]$ are now considered.
As it is shown below, these limits do not commute.
The limits $y_o\rightarrow\pm\infty$ are first evaluated in the following
theorem.

\begin{theorem} \label{th:paquetremarque*}
If the wave function $\wave$ is defined by eq.~(\ref{eq:packet}), then for 
all $\delta>0$
\be
\lim_{y_o \rightarrow +\infty} 
\left(\,\bar{N}\,[\,\pa\,]\ -\bar{N}^{Th}_{2\pi}[\,\pa\,]\,\right) &=& 0,
\\ [2mm]
\lim_{y_o \rightarrow -\infty} \bar{N}[\,\pa\,] &=& 0,
\ee
and thus
\be
\lim_{\delta\rightarrow0}\ \lim_{y_o \rightarrow +\infty} 
\left(\,\bar{N}\,[\,\pa\,]\ -\bar{N}^{Th}_{2\pi}[\,\pa\,]\,\right) &=& 0,
\\ [2mm]
\lim_{\delta\rightarrow0}\ \lim_{y_o \rightarrow -\infty} \bar{N}[\,\pa\,] 
&=& 0.
\ee
\end{theorem}
\paragraph{Proof}
Since the function $\wave(y)$ decreases at least as $1/\vert y\vert^2$ at 
infinity and property $v)$ is satisfied by assumption, theorems 
\ref{th:approche} and \ref{th:flat} can be applied to this wave function and 
one concludes immediately.
\hfill{$\Box$}

The generalised function $\packets$ is now defined by
\be
\packets(p) &=& \lim_{\gd\rightarrow0^+} \packet(p), \qquad \forall\,p\in\R.
\label{eq:packets}
\ee
The expectation value of the momentum operator in the corresponding state
equals $p_o$ and its variance vanishes in this state.  Notice that, in 
configuration space, the function $\pa$ becomes more and more
extended in the limit $\gd \rightarrow 0^+ $, although, if
$\packet\in\cC^\infty(\R)$, $\pa$ decreases at infinity faster than the 
inverse of any algebraic function of $y$.  Theorems \ref{th:approche} and 
\ref{th:flat} may thus not be applied to $f=\waves$, because the quantity $L$ 
defined in eqs (\ref{eq:dalgebrique}) or (\ref{eq:dalgebriqueplat}) tends to 
infinity when $\gd\rightarrow0^+$, and consequently the bounds 
(\ref{eq:approche}) and (\ref{eq:plat}) diverge exponentially in this limit. 
The following lemma will be needed to consider the physics of the delocalised 
wave function $\waves$.

\begin{lemma} \label{l:packet}
Let $\gd$ and $\gep$ be two positive constants.  If $\gd$ is small 
enough, one has for all $y_o\in(-\gep/\gd,\gep/\gd)$
\be  
\left\vert\,\bar{N}\,[\,\pa\,]-
\sur{1}{2}\bar{N}^{Th}_{2\pi}\,[\,\pa\,]\,\right\vert
&\leq& C\,\left[\,n(p_o)\,\gd +\sur{\gep(\gep+1)}{e^{2\pi p_o}-1}\,\right],
\label{eq:npaquet}
\ee
where $C>0$ is a constant and where the function $n$ satisfies
\be
\abs{n(p_o)} &\leq& \left\{
\begin{array}{cl}
\sur{\log^2 p_o}{p_o}, \quad & \mbox{if \ $p_o \approx 0^+$,} \\ [4mm]
p_o^4 \, e^{-4\pi p_o}, \quad & \mbox{if \ $p_o \gg 1$,}
\end{array} \right.
\label{eq:paquetbornendelta}
\ee
and is bounded except in the neighborhood of $p_o=0$.
\end{lemma} 
\paragraph{Proof} 
See appendix \ref{ap:packet}.
\\ \\
This lemma implies that the mean number $\bar{N}\,[\,\pa\,]$ is approximately
equal to the {\it half} of the thermal average 
$\bar{N}^{Th}_{2\pi}\,[\,\pa\,]$ if $\gd$ and $\gep$ are small enough. It may 
be applied to the generalised function $\waves$ to calculate
$\bar{N}\,[\,\wapas\,]$.

\begin{theorem} \label{th:paquetremarque}
If the generalised function $\waves$ is defined by eq.~(\ref{eq:packets}), then
\be
\bar{N}\,[\,\wapas\,] &=& \sur{1}{2}\,\sur{1}{e^{2\pi p_o}-1},
\saut \forall\,y_o\in\R,
\label{eq:paquetthermique}
\ee
where $p_o>0$, and in consequence
\be
\lim_{y_o \rightarrow \pm\infty} \ \lim_{\delta\rightarrow0}\, \left(\,
\bar{N}\,[\,\pa\,]-\sur{1}{2}\,\bar{N}^{Th}_{2\pi}\,[\,\pa\,]\,\right) &=& 0,
\label{eq:paquetremarquableA}
\ee
where the wave function $\wave$ is defined in eq.~(\ref{eq:packet}).
\end{theorem}
\paragraph{Proof}
The limit $\gd\rightarrow0^+$ is first evaluated in eq.~(\ref{eq:npaquet}) of 
lemma \ref{l:packet} and the result obtained is then true for all 
$y_o\in\R$.  The limit $\gep\rightarrow0^+$ is next evaluated and 
eq.~(\ref{eq:paquetthermique}) is obtained. Equation 
(\ref{eq:paquetremarquableA}) follows then from eq.~(\ref{Nthermal}).
\hfill{$\Box$} \\ \\
This theorem shows again that the mean number of particles created in a state
may not be thermal close to the horizon and may not vanish far from the black
hole.  Theorems  \ref{th:paquetremarque*} and \ref{th:paquetremarque}
imply that the limits $\delta \rightarrow 0^+$ and $y_o\rightarrow \pm \infty$ 
of the mean number $\bar{N}\,[\,\pa\,]$ do not commute.

\section{Conclusions}

In the present paper, exact calculations of the mean number $\bar{N}[f]$ of 
massless scalar particles created spontaneously in a given state $f$ have 
been performed in the CGHS black-hole background. Since our approach do not 
rely on the approximation of the horizon, one was able to draw some rigorous
conclusions on the issues related to the convergence of the mean number
$\bar{N}[f]$ and to the approach to the thermal equilibrium as well, 
and to calculate exactly the mean number of particles created in a given mode.

The main conclusion of this paper is that the physics close - or 
asymptotically close - to the horizon depends on the global properties of 
the considered state.  For example, the approach to the thermal equilibrium 
of the mean number $\bar{N}[f]$ close to the horizon depends on the queue 
of the state far from the black hole, if this decreases sufficiently fast, 
otherwise $\bar{N}[f]$ may be not thermal in that limit.
Similarly, the mean number of particles created in a given mode is not thermal
even close to the horizon, because the corresponding state is unlocalised,
and the contribution of the part of the state which is far from the black hole
must also be taken into account. Since $\bar{N}[f]$ is essentially not a local 
quantity, and because a state without negative momentum components cannot be 
localised, the mean number of particles $\bar{N}[f]$ created close to the 
horizon depends to some extent on the spacetime properties far from the 
black hole.

\subsection*{Acknowledgements}

I thank G. Wanders for support, stimulating discussions and criticism.

\appendix
\section{Appendices}

\subsection{Two useful theorems} \label{sub:apth}

\begin{theoremA}[Paley-Wiener \cite{Ch}] \label{th:causal}
Assume that $g\in\LLd{y}$.  Then there is a function 
$\go: \R \rightarrow \R$ such that the Fourier transform of 
$f(y)\equiv g(y)\, e^{\I \go(y)}$ satisfies 
\be
\tf{f}(p) &=& \gt(p) \, \tf{f}(p), \saut \forall\,p\in \R,
\label{eq:paleywiener*}
\ee
if and and only if
\be
\left\vert \, \int_{-\infty}^{+\infty} dy \
\sur{\log \abs{g(y)}}{1+y^2} \ \right\vert &<& \infty.
\label{eq:paleywiener}
\ee
In particular, eq.~(\ref{eq:paleywiener*}) implies that 
the modulus $\abs{f}$ is strictly bounded at infinity from below by a 
decreasing exponential function.
\end{theoremA}

\begin{theoremA}[Hilbert transform \cite{Ch}] \label{th:hilbert}
If $\tf{g} \in \LLd{p}$, then
\be
\tf{f}(p) &\equiv& \frac{1}{\pi} \, P \int_{-\infty}^{+\infty} dp' \, 
\sur{\tf{g}(p')}{p'-p}
\ee
converges almost everywhere if $p\in\R$. Furthermore, one has 
$\tf{f} \in \LLd{p}$ and
\be
\tf{g}(p) &=& -\frac{1}{\pi} \, 
P \int_{-\infty}^{+\infty} dp'\,\sur{\tf{f}(p')}{p'-p}, 
\qquad \forall\,p\in\R, \\ [2mm]
\int_{-\infty}^{+\infty} dp\, 
\abs{\tf{g}(p)}^2 &=& \int_{-\infty}^{+\infty}dp\,\abs{\tf{f}(p)}^2.
\ee
\end{theoremA}

\subsection{Proof of lemma \ref{lemma:f*}} \label{sub:apf*}

\begin{lemmaA} \label{lemmaA:f*}
Let be a differentiable function $g\in\LLup{x}$ such that $\tf{g}(0)=0$, and
define the function $h$ by
\be
g(x) &=&  \sur{h(x)}{x \, (-\log x)^{1+\ga}}, \qquad \mbox{if \ $x>0$,}
\label{eq:Af*Defh}
\ee 
where $\ga > 0$.  If the limit $C\in\C$ of $h$ exists when 
$x\rightarrow +\infty$ (or $x\rightarrow 0^+$), and if one has 
\be
h'(x) &=& \OO \left(\sur{1}{x\,\log x}\right)
\label{eq:Af*Hyp}
\ee  
when $x\gg1$ (or $x\approx 0^+$), then
\be 
\tf{g}(-k) &=& \sur{C}{\sqrt{2\pi}}\,\sur{A_g(k)}{\ga\,\left(\log k\right)^\ga}
+ \OO \left[\sur{1}{(\log k)^{\ga+1}}\right],
\label{eq:Af*}
\ee
where $A_g$ is a function depending on $g$ and satisfying
\beaa{rcccl}
1/2 &\leq& \abs{A_g(k)} &\leq& 3/2
\label{AIneq}
\eeaa
when $k\approx 0^+$ (or $k\gg 1$).
\end{lemmaA}

\paragraph{Proof}
The proof will be only sketched here (see ref.~\cite{thesis} for a complete
proof).
The idea is to split the Fourier transform $\tf{g}(-k)$ into two terms: 
\be
\sqrt{2\pi} \, \tf{g}(-k) &=& \int_0^{\frac{1}{2k}} dx \, g(x) \, e^{\I kx} +
\int_{\frac{1}{2k}}^\infty dx \, g(x) \, e^{\I kx}.
\ee
Under assumption (\ref{eq:Af*Hyp}), one shows that the second term on the
r.h.s.~of this last equation is of higher order than the first one by 
integrating by part. One next defines the function $A_g(k)$ by
\be
\int_0^{\frac{1}{2k}}dx\,g(x)\,e^{\I kx} &=& 
A_g(k)\,\int_0^{\frac{1}{2k}}dx\,g(x),
\label{eq:Af*a}
\ee 
from which inequalities (\ref{AIneq}) are deduced from the L'Hospital
rule.  Approximating and integrating the r.h.s.~of eq.~(\ref{eq:Af*a})
yields eq.~(\ref{eq:Af*}) from eqs (\ref{eq:Af*Defh}) and (\ref{eq:Af*Hyp}).
\hfill{$\Box$}

\begin{lemmaA} \label{lemmaA:f*bis}
If $g\in\LLup{x}$ is such that $\tf{g}(0)=0$, and if its modulus
$\vert g\vert$ satisfies
\be
\abs{g(x)}&\leq& \sur{2\,C}{(x-\log2)^{1+\gep}}, \qquad\mbox{if \ $x\geq l$,}
\label{lemmaA:f*bisHyp}
\ee
where $C>0$, $l>\log2$, and $\gep>0$ are three constants, then 
\be
\sqrt{2\pi}\,\abs{\tf{g}(-k)} &\leq&
\sur{4\,C}{q(\gep)}\,k^{q(\gep)}\,
\left[\,e^{q(\gep)}+\sur{1}{2\,[\,1-q(\gep)\,]} \,\right]
+ k\,(e^l-1)\,\no{g}_{\LLup{x}},
\label{lemmaA:f*bisRes}
\ee
when $0\leq k\leq l^{-1}$, and where the function $q$ is defined in 
eq.~(\ref{defq}).
\end{lemmaA}

\paragraph{Proof}
The primitive $G$ of $g$ is defined by $G(x)=\int_0^x dx'\,g(x')$.
Since $\tf{g}(0)=G(0)=0$, one gets by splitting the integral and by 
integrating by parts,
\be
\sqrt{2\pi} \, \tf{g}(-k) 
&=& -\I\,k \int_0^{1/k} dx\,G(x)\ e^{\I kx}
+\ \int_{1/k}^\infty dx\,g(x)\,\left(\,e^{\I kx}-e^\I\,\right).
\label{eq:prelilemmeCA}
\ee
From assumption (\ref{lemmaA:f*bisHyp}), a bound for the first term on the
r.h.s.~of this last equation is obtained by
writing $\int_0^{1/k}dx=\int_0^ldx+\int_l^{1/k}dx$, where 
$0\leq k\leq l^{-1}$, and by noting that the behavior of the bound for 
$\int_l^{1/k}dx\, G(x)\ e^{\I kx}$ depends on $\gep$. 
A bound for the second term is easily obtained from assumption 
(\ref{lemmaA:f*bisHyp}) as well and one gets finally
eq.~(\ref{lemmaA:f*bisRes}).
\hfill{$\Box$}

\paragraph{Proof of lemma \ref{lemma:f*}}
From eqs (\ref{decreasef}) and (\ref{decreasefbis}), one has
\be
\abs{\ind{\ch{f}}{\ga}(-x)} &\leq& 
\sur{2\,\vert C_{\ga,-}\vert}{(x-\log2)^{1+\ga}}
+\OO\left(\sur{1}{x^{2+\ga}}\right),  \hspace{19mm} \mbox{{\it if} \ $x\gg 1$,}
\label{eq:lemmaAf*bis} \\ [2mm]
\ind{\ch{f}}{\ga}(-x) &=& 
\sur{C_{\ga,+}}{x\, (-\log x)^{1+\ga}}
+\OO\left[\sur{1}{x\, (-\log x)^{2+\ga}}\right], 
\quad \mbox{{\it if} \ $x\approx 0^+$,} 
\label{eq:lemmaAf*} 
\ee
and
\be
\Indice{\ch{f}}{\ Th}{\ga}(-x)  
&=&  \sur{C_{\ga,\mp}}{x\, (-\log x)^{1+\ga}}
+\OO\left[\sur{1}{x\, (-\log x)^{2+\ga}}\right], \quad 
\mbox{{\it if}\ } \left\{ \begin{array}{l}
x \gg 1, \\ [4mm]
x\approx 0^+.
\end{array} \right.
\ee
To obtain eq.~(\ref{eq:lemmaAf*bis}), one made use of
$\vert y(-x)\vert\,\geq x-\log2$ and $y'(-x)\leq2$ if $x\geq\log2$.
Equations (\ref{eq:infiniTN}), (\ref{eq:zerohorizon}) and 
(\ref{eq:infinihorizon}) follow then from lemma \ref{lemmaA:f*} if 
$g(x)=\ind{\ch{f}}{\ga}(-x)$.  Equation (\ref{eq:zeroTN}) follows from lemma 
\ref{lemmaA:f*bis}.
\hfill{$\Box$}

\subsection{Proof of lemma \ref{lemma:f**}} \label{sub:apf**}

\begin{lemmaA} \label{lemmaA:f**}
If $g\in\LLup{x}$ is a function such that $\tf{g}(0)=0$, and if its
modulus $\vert g\vert$ satisfies
\be 
\abs{g(x)} &\leq& \sur{C_-}{x\,(\log x)^{1+\ga}},
\saut \mbox{if \ $x\geq l$,}
\label{eq:prelihypoA}
\ee  
where $C_->0$, $l>1$, and $\ga>0$ are three constants, then
\be
\sqrt{2\pi} \,\abs{\tf{g}(-k)} &\leq&  
\sur{3\,C_-}{\ga\,(-\log k)^\ga} + k\,l\,\no{g}_{\LLup{x}}
\label{eq:prelilemmeA}
\ee 
where $0\leq k\leq l^{-1}$.
\end{lemmaA}

\paragraph{Proof}
The proof is similar to the one of lemma \ref{lemmaA:f*bis}. In this case, 
however, the behavior of the bound for $\int_l^{1/k}dx\,G(x)\ e^{\I kx}$ one
obtains does not depend on the parameter $\ga$. 
\hfill{$\Box$}

\begin{lemmaA} \label{lemmaA:f**bis}
If $g\in\LLup{x}$ is a differentiable function such that its modulus satisfies
\be 
\abs{g(x)} &\leq& \sur{C_+}{x\,(-\log x)^{1+\ga}},
\qquad \mbox{if \  $0<x\leq l$,}
\label{eq:prelihypoB}
\ee   
where $C_+>0$, $0<l<1$ and $\ga>0$ are three constants, then
\be
\sqrt{2\pi} \,\abs{\tf{g}(-k)} &\leq& 
\sur{2^\ga\,C_+}{\ga\,(\log k)^\ga}
+ \sur{1}{\sqrt{k}}\,\sur{2^{1+\ga}\,C_+}{(\log k)^{1+\ga}}
+\sur{1}{k}\,\int^\infty_{\frac{1}{\sqrt{k}}} dx\,\abs{g'(x)} 
\label{eq:prelilemmeB}
\ee
where $k\geq l^{-2}$.
\end{lemmaA} 

\paragraph{Proof}
The Fourier transform $\tf{g}(k)$ is split into two terms,
\be
\sqrt{2\pi} \, \tf{g}(-k) &=&
\int_0^{\frac{1}{\sqrt{k}}} dx \, g(x) \, e^{\I kx} 
+\int_{\frac{1}{\sqrt{k}}}^\infty dx \, g(x) \, e^{\I kx}. 
\label{eq:prelilemmetfbis}
\ee
A bound for the first term on the r.h.s.~of this last equation is obtained 
directly from assumption (\ref{eq:prelihypoB}) if $k\geq l^{-2}$, and a bound 
for the second term is deduced by integrating it by parts. Equation
(\ref{eq:prelilemmeB}) then follows.~\hfill{$\Box$}

\paragraph{Proof of lemma \ref{lemma:f**}}
From eqs (\ref{eq:lemmaf**}) and (\ref{eq:lemmaf**bis}), one has
\be
\abs{\ch{f}(-x)} &\leq& 
\left\{ \begin{array}{ll}
\sur{C_-}{x^{1+\gep}}, & \mbox{{\it if} \ $x\gg 1$,}
\label{eq:lemmaAf**bis} \\ [4mm]
\sur{C_+}{x\, (-\log x)^{1+\ga}},
\quad & \mbox{{\it if} \ $x\approx 0^+$,} 
\label{eq:lemmaAf**} 
\end{array}\right.
\ee
and
\be
\abs{\indsup{\ch{f}}{\,Th}(-x)} &\leq& 
\left\{ \begin{array}{ll}
\sur{C_-}{x\, (-\log x)^{1+\gep}},
&\quad \mbox{{\it if} \ $x\gg 1$,} \\ [4mm]
\sur{C_+}{x\, (-\log x)^{1+\ga}},
&\quad \mbox{{\it if} \ $x\approx 0^+$.} 
\end{array}\right.
\ee
Lemmas \ref{lemmaA:f*bis} and \ref{lemmaA:f**} imply
eqs (\ref{lemmaf**A}) and (\ref{lemmaf**C}) respectively.
Lemma \ref{lemmaA:f**bis} is next applied to $g(x)=\ch{f}(-x)$. Since
\be
x'(y) \left.\frac{d}{dx}\ch{f}(x)\right\vert_{x(y)}
&=& -\sur{x''(y)}{x'(y)^2}\,f(y)+\sur{1}{x'(y)}\,f'(y),
\label{eq:compactthermformule}
\ee
one has
\be
\Int_{\frac{1}{\sqrt{k}}}^\infty dx
\,\left\vert\,\frac{d}{dx}\ch{f}(-x) \,\right\vert \ 
&\leq& \Int^{\log \sqrt{k}}_{-\infty} \,dy\ \left(\,1+e^y\,\right)\,
\left(\,\abs{f(y)}+ \abs{f'(y)}\,\right).
\ee
This result is also true for $g(x)=\indsup{\ch{f}}{\,Th}(-x)$,
and eqs (\ref{lemmaf**B}) and (\ref{lemmaf**D}) are then obtained.
\hfill{$\Box$}

\subsection{Proof of lemma \ref{l:bound}} \label{sub:aplbound}

From eqs (\ref{eq:trounoirnoyauU}) and (\ref{defN}) one has
\be
\bar{N}[f] \ = \ \sur{1}{8\pi^3}
\Int_0^\infty dp' \, \tf{f}(p')^* \Int_0^\infty dp \, \tf{f}(p)\,
\hspace{80mm} \label{eq:trounoirborneA} \\ [1mm] \times \ 
\gC(\I p') \, \gC(-\I p) \ 
\Int_0^\infty dk\,\sinh(\pi k)\,\gC(-\I p'-\I k+0^+)\,\gC(\I p+\I k+0^+),
\hspace{5mm} \nonumber
\ee
where eq.~(\ref{eq:identitegamma}) has been used.
Stirling's formula \cite{GuW},
\be
\gC(z) &=& \sqrt{\sur{2\pi}{z}} \, e^{-z} \, z^z 
\left[ \, 1+ {\cal O}\left(\frac{1}{z}\right) \, \right],
\ee
where $\abs{\arg z}\ < \pi$, implies that
\beaa{l}
\sinh (\pi k) \, \gC(-\I p'-\I k+0^+) \, \gC(\I p+\I k+0^+)
\hspace{70mm} \\ [4mm] \hspace{50mm}
= \ \pi \, e^{-\frac{\pi}{2} (p+p')}\, \sur{e^{\I (p-p')\log k}}{k} \,
\left[ \, 1+ {\cal O}\left(\sur{1+p^2+{p'}^2}{k}\right) \, \right] 
\cdot
\label{eq:spontaneTNborneexpressionA}
\eeaa
The integral over $k$ in eq.~(\ref{eq:trounoirborneA}) is split into three 
terms according to the partition $\R_+ = [0,w]\cup(w,W)\cup [W,\infty)$, 
where $w$ and $W$ are two positive constants which are small and large enough 
respectively.
The contribution of the non-compact interval $[W,\infty)$ is given from 
eq.~(\ref{eq:spontaneTNborneexpressionA}) by
\beaa{l}
\Int_W^\infty \sur{dk}{2k} \, U(-k,p')^* \, U(-k,p) 
\ = \ \sur{\delta(p-p')}{4p\,(e^{2\pi p}-1)} \ + 
\hspace{70mm} \\ [2mm] \hspace{15mm}
+ \ \sur{\I}{8\pi^2} \, t(p) \, t(p')^* P\, \sur{1}{p-p'}
+\sur{1}{\sqrt{p\,(e^{2\pi p}-1)\,p'\,(e^{2\pi p'}-1})} \, 
{\cal O} \left(\sur{1+p^2+{p'}^2}{W}\right),
\label{eq:spontaneTNborneexpressionB}
\eeaa
where the function $t$ has been defined by
\be
t(p) &=& \gC (-\I p) \ e^{-\frac{\pi}{2} p} \ e^{\I p\log W},
\ee
and satisfies eq.~(\ref{eq:t(p)}) (see eq.~(\ref{eq:identitegamma})).
The first term on the r.h.s.~of eq.~(\ref{eq:spontaneTNborneexpressionB}) is 
the half of the kernel of $\bar{N}^{Th}_{2\pi}[f]$ (see eq.~(\ref{Nthermal})),
and thus
\be
\Int_0^\infty dp'\,\tf{f}(p')^*\Int_0^\infty dp\,\tf{f}(p)
\Int_W^\infty \sur{dk}{2k} \, U(-k,p')^* \, U(-k,p) 
\ = \ \sur{1}{2} \bar{N}^{Th}_{2\pi}[f]  
\hspace{35mm} \\ [-8mm] \nonumber
\ee
\be
+ \ \sur{\I}{8\pi^2} \Int_0^\infty dp' \tf{f}(p')^* \, t(p')^*
        \Int_0^\infty dp \, P \sur{\tf{f}(p) \, t(p)}{p-p'} \ 
+ \ {\cal O}\left(\sur{1}{W}\right) \left\vert \,\Int_0^\infty dp\, 
      \sur{\tf{f}(p)\, (1+p^2)}{\sqrt{p(e^{2\pi p}-1)}} \, 
\right\vert^2. \nonumber
\ee
If  $\gep>0$ is small enough one has furthermore
\beaa{l}
\left\vert\,\Int_0^w dk\,\sinh(\pi k)\,\gC(-\I p'-\I k+0^+)\,\gC(\I p+\I k+0^+)
\,\right\vert \hspace{60mm} \\ [4mm] \hspace{55mm}
\leq \ C\, e^{-\frac{\pi}{2}(p+p')} \ \times 
\left\{ \begin{array}{l}
\log\left(1+\sur{w}{\gep}\right),
\mbox{\ \ {\it if} \ $p,p'\geq\gep>0$,} \\ [3mm]
\log\left(1+\sur{w}{p}\right), \mbox{\ \ {\it otherwise,}}
\end{array} \right. \\ [-4mm]
\eeaa
\beaa{l}
\left\vert \, \Int_w^W dk \,
\sinh (\pi k) \, \gC(-\I p'-\I k+0^+) \, \gC(\I p+\I k+0^+) \, 
\right\vert \ \leq \
C\,\sur{W}{w} \, e^{-\frac{\pi}{2}(p+p')}, \hspace{25mm}
\eeaa
where $C>0$ is a constant.
From the last three equations one gets
\be
\left\vert\,\bar{N}[f]-\sur{1}{2}\bar{N}^{Th}_{2\pi}[f]\,\right\vert &\leq&
\sur{1}{8\pi^2}\left\vert\,\Int_0^\infty dp'\,\tf{f}(p')^*\,t(p')^*
\Int_0^\infty dp\ P\,\sur{\tf{f}(p)\,t(p)}{p-p'}\,\right\vert
\hspace{40mm} \label{eq:Bornetn} \\ [-8mm] \nonumber
\ee
\be
+\ C\,\left[ \, \Int_0^\infty dp \,
  \sur{\abs{\tf{f}(p)}}{\sqrt{p(e^{2\pi p}-1)}}\,(1+p^2)\,\right]^2
+\ C\,\left[ \, 1+\log \left(1+\sur{w}{\gep}\right)\,\right] 
  \left[\,\Int_\gep^\infty dp\, 
  \sur{\abs{\tf{f}(p)}}{\sqrt{p(e^{2\pi p}-1)}}\,\right]^2 
\nonumber \\ [1mm] \nonumber
+ \ C\,\Int_0^\gep dp \, 
  \sur{\abs{\tf{f}(p)}}{\sqrt{p(e^{2\pi p}-1)}}\,\log\left(1+\sur{w}{p}\right)
  \Int_0^\infty dp \,\sur{\abs{\tf{f}(p)}}{\sqrt{p(e^{2\pi p}-1)}}
\ee
which implies eq.~(\ref{eq:bound}).
\hfill{$\Box$}

\subsection{Proof of lemma \ref{lemma:AN*}} \label{ap:AN*}

The Fourier transform of $\fy{\ch{f}}(x)$ is split into two terms,
\be
\intio dx \, \fy{\ch{f}}(x) \, e^{\I kx}  &=&
\Int_0^{\frac{1}{\sqrt{kx_o}}} dx \, \fy{\ch{f}}(-x) \, 
\left(e^{-\I kx}-1\right)
+ \Int_{\frac{1}{\sqrt{kx_o}}}^\infty dx \, \fy{\ch{f}}(-x) \,
\left(e^{-\I kx}-1\right) \ \ \ \ \ \
\label{eq:ineaA}
\ee
where $x_o \equiv e^{y_o}$.  A bound for the first term on the r.h.s.~of this
last equation is given for both the CGHS black-hole and thermal cases by
\be
\left\vert \, \int_0^{\frac{1}{\sqrt{kx_o}}} 
dx \, \fy{\ch{f}}(-x) \,\left(e^{-\I kx}-1\right)\,\right\vert
&\leq & \sqrt{k} \, e^{-y_o/2} \, \Lu{f}.
\label{eq:ineaB}
\ee 
To obtained a bound for the second term, the CGHS black-hole and thermal cases
must be treated separately, although similar bounds will be found for these
two cases.  One assumes in both cases that $e^{L-y_o} \leq x$, so that use of 
assumption (\ref{eq:dalgebrique}) can be made.  In the second term one has 
$(kx_o)^{-0.5}\leq x$, one must thus have $e^{L-y_o}\leq (kx_o)^{-0.5}$. This 
means that the bounds obtained below are true for $0\leq k\leq e^{y_o-2L}$.

The thermal case is first considered.  Assumption (\ref{eq:dalgebrique*}) 
implies
\be
\abs{\Indice{\ch{f}}{\,Th}{y_o}(-x)}
&\leq& \sur{C}{x \, \left[\,\log (xx_o) \, \right]^{1+\ga}},
\qquad \mbox{{\it if} \ $x \geq e^{L-y_o}$,}
\label{eq:ineaC}
\ee
from which one deduces, if $0\leq k\leq e^{y_o-2L}$, that
\be
\left\vert \, \Int_{\frac{1}{\sqrt{kx_o}}}^\infty dx \, 
\Indice{\ch{f}}{\,Th}{y_o}(-x) \, 
\left(e^{-\I kx}-1\right) \, \right\vert
&\leq & \sur{2C}{\ga}\,\sur{2^\ga}{\left(\,y_o-\log k\, \right)^{\ga}}
\cdot
\label{eq:ineaCbis}
\ee
Equations (\ref{eq:ineaB}) and (\ref{eq:ineaCbis}) imply eq.~(\ref{eq:inea}) 
for the thermal case.

The case of the CGHS black hole is now considered.
An auxiliary transformation $\ys(x)$ is defined by
\be
\ys(x) &=& \left\{ 
\begin{array}{cl}
-\log x, \quad & \mbox{{\it if} \ $0<x<\log 2$,} \\ [4mm]
-x + \log 2, \quad & \mbox{{\it if} \ $\log 2\leq x< \infty$.}
\end{array}
\right.
\ee
Since
\be
y'(x) \ = \ \sur{1}{1-e^{-x}} &\leq& 2 \times \left\{
\begin{array}{cl}
1/x, \quad& \mbox{{\it if} \ $0<x<\log 2$,} \\ [4mm]
1, & \mbox{{\it if} \ $\log 2\leq x< \infty$,}
\end{array}
\right.
\ee
assumption (\ref{eq:dalgebrique*}) implies when $x\geq e^{L-y_o}$ that
\be
\abs{\ind{\ch{f}}{y_o}(-x)} &\leq& 
\sur{2\,C}{\left[\,y_o -\ys(x)\, \right]^{1+\ga}} \times \left\{ 
\begin{array}{cl}
1/x, \quad & \mbox{{\it if} \ $0<x<\log 2$,} \\ [4mm]
1, & \mbox{{\it if} \ $\log 2\leq x< \infty$.}
\end{array} \right.
\label{eq:ineaD}
\ee
The cases $(kx_o)^{-0.5}\leq \log 2$ and $(kx_o)^{-0.5}\geq\log 2 $ have to
be considered separately.  If $(kx_o)^{-0.5}\leq \log 2$ , one has from 
eq.~(\ref{eq:ineaD})
\be
\Int_{\frac{1}{\sqrt{kx_o}}}^\infty dx \, 
\abs{\ind{\ch{f}}{y_o}(-x)}
&\leq&  \Int_{\frac{1}{\sqrt{kx_o}}}^{\log 2} dx \, 
\sur{2\,C}{x \, \left[\,\log (xx_o) \, \right]^{1+\ga}}
+ \Int_{\log 2}^\infty dx \, \sur{2\,C}{(\,x+y_o-\log 2\,)^{1+\ga}}
\nonumber \hspace{10mm} \label{eq:ineaE} \\ [3mm] &\leq&
\sur{2\,C}{\ga}\,\sur{2^\ga}{\left( \,y_o-\log k\, \right)^{\ga}}\cdot  
\label{eq:ineaF}
\ee
If $(kx_o)^{-0.5}\geq\log2$, one has also from eq.~(\ref{eq:ineaD})
\be
\Int_{\frac{1}{\sqrt{kx_o}}}^\infty dx \, 
\abs{\ind{\ch{f}}{y_o}(-x)} \ 
\ \leq \ \Int_{\frac{1}{\sqrt{kx_o}}}^\infty dx \, 
\sur{2\,C}{(\,x+y_o-\log 2\,)^{1+\ga}}
\ \leq\ \sur{2\,C}{\ga}\,\sur{2^\ga}{\left(\,y_o-\log k-1\,\right)^{\ga}},
\label{eq:ineaG} \label{eq:ineaI}
\ee
since $0<-2\log z\leq1/z$ (if $0<z<1$) implies for $z=\log2\,\sqrt{kx_o}$
\be
0< - \log 2 \, (\,2 \log \log 2 + \log k + y_0) 
\ \leq \sur{1}{\sqrt{kx_o}}\cdot
\label{eq:ineaH}
\ee 
Equation (\ref{eq:inea}) for the black hole case is then deduced from eqs 
(\ref{eq:ineaB}), (\ref{eq:ineaF}) and (\ref{eq:ineaI}).~\hfill{$\Box$}

\subsection{Proof of lemma \ref{lemma:AN**}} \label{ap:AN**}

The subscript $y_o$ is dropped in this subsection for clarity. Defining the
transformation $\xh=\xh(y)$ by eq.~(\ref{x(y)gui}), one gets by integrating 
by parts
\be
\intio dx \, \gd \ch{f}(x) \, e^{\I kx} = 
\sur{\I}{k}\intii dy\,\left[\,x'(y)\,\ch{f}'(x(y))\ e^{\I kx(y)}
- \xph(y)\, \indsup{\ch{f}'}{Th} (\xh(y)) \ e^{\I k\xh(y)} 
\,\right] \ \ \
\ee
Equation (\ref{eq:compactthermformule}) implies
\be
\intio dx \, \gd \ch{f}(x) \, e^{\I kx} &=&
\sur{\I}{k}\left\{\,-\intii dy\,f'(y)\left[\,
\sur{1}{x'(y)}\,e^{\I kx(y)}-\sur{1}{\xph(y)}\,e^{\I k\xh(y)}
\,\right]\right. \hspace{25mm} \nonumber \\ [2mm] && \left.
+\intii dy\,f(y)\left[\,\sur{x''(y)}{x'(y)^2}\,e^{\I kx(y)}
-\sur{\xpph(y)}{\xph(y)^2}\,e^{\I k\xh(y)}\,\right]\ \right\}
\nonumber \\ [-4mm] \nonumber
\ee
\be
\ =\ \sur{\I}{k} \left\{\,  
\intii dy \, e^y \, \left[ \, f(y)+f'(y)\, \right] \,
\left[ \, e^{\I kx(y)}-e^{\I k\xh(y)} \, \right]+
\intii dy \, f'(y)\, e^{\I kx(y)} \,\right\}.
\hspace{15mm}\label{eq:ineqbB}
\ee
Now if one defines $z=e^{-y}$ one gets
\be
e^y \, \left\vert \, e^{-\I kx(y)}-e^{-\I k\xh(y)}\, \right\vert &\leq & 
\left\{ \begin{array}{c}
\sur{k}{z}\,\left[\,z-\log (1+z)\,\right], \\ [3mm]
\sur{2}{z},
\end{array} \right.
\ee
and in particular
\be
e^y \, \left\vert \, e^{-\I kx(y)}-e^{-\I k\xh(y)}\, \right\vert
&\leq & \sur{2}{z_m},
\label{eq:inecA}
\ee
where $z_m$ is defined by $k\,\left[\,z_m-\log(1+z_m)\,\right]=2$.
Since $0\leq z- \log (1+z) \leq \sur{z^2}{2}$, one obtains 
$\sur{2}{z_m}\leq\sqrt{k}$ and thus
\be
e^y\,\left\vert\,e^{\I kx(y)}-e^{\I k\xh(y)}\,\right\vert
&\leq& \sqrt{k}.
\label{eq:inec}
\ee
Equations (\ref{eq:ineqbB}) and (\ref{eq:inec}) imply finally
eq.~(\ref{eq:ineb}).
\hfill{$\Box$}

\subsection{Proof of lemma \ref{l:flat*}} \label{ap:flat*}

\paragraph{a)}
Equation (\ref{eq:dalgebriqueplat*}) implies eq.~(\ref{lemmaA:f*bisHyp}) with 
$l=\log(1+e^L)\leq e^L$ since $\vert y(-x)\vert\,\geq x-\log2$ and 
$y'(-x)\leq2$ if $x\geq\log2$.  Lemma \ref{lemmaA:f*bis} is then applied 
with $\ga>1/2$ and eq.~(\ref{eq:flat*A}) is obtained.

\paragraph{b)}
Since $\tf{f}(-k)=0$ if $k\geq0$, one has
\be
\sqrt{2\pi}\,\tfch{f}(-k) &=& \intii dy\,f(y)
\,\left[\,e^{\I kx(y)}-e^{\I ky}\,\right].
\ee
One writes then
\be
\sqrt{2\pi}\,\ \ind{\tfch{f}}{y_o}(-k) \ = \hspace{120mm}
\nonumber \\ [2mm] 
\int_{-\infty}^{L+y_o/2}dy\,f_{y_o}(y)\,e^{\I ky}\,
\left\{\,e^{\I k\,[\,x(y)-y\,]}-1\right\} 
+\int^{+\infty}_{L+y_o/2}dy\,f_{y_o}(y)\,
\left[\,e^{\I kx(y)}-e^{\I ky}\,\right]. \hspace{10mm}
\label{eq:l:flat*}
\ee
A bound for the first integral on the r.h.s.~of this last equation is 
obtained from
\be
\vert\,x(y)-y\,\vert &\leq& e^y,\qquad \forall\,y\in\R.
\ee
Assumption (\ref{eq:dalgebriqueplat*}) is used in the second integral 
to deduce eq.~(\ref{eq:flat*B}). 

\paragraph{c)}
Equation (\ref{eq:flat*C}) is obtained in the same way as 
eq.~(\ref{lemmaf**B}) of lemma \ref{lemma:f**}.
\hfill{$\Box$}

\subsection{Proof of lemma \ref{l:packet}} \label{ap:packet}

This lemma is deduced from eq.~(\ref{eq:bound}) of lemma \ref{l:bound}.
The second term on the r.h.s.~of this equation gives a contribution of order 
$\gd$ if $f=\pa$.
To treat the first term with the principal value, one writes in 
eq.~(\ref{eq:packet*}),
\be
e^{-\I py_o} &=& e^{-\I p_oy_o}+e^{-\I p_oy_o}\,
\left[\,e^{-\I (p-p_o)\,y_o}-1\,\right].
\ee
The real contribution of $e^{-\I p_oy_o}$ in this first term vanishes
because the function $\gD_\gd$ is real by assumption and the integrand is 
anti-hermitian.
The imaginary contribution of $e^{-\I p_oy_o}$ in the first term is 
proportional to
\be
\intoi dp\intoi dp' \abs{ \tfpa(p)\,t(p)\,\tfpa(p')\,t(p') }
\,\sur{\sin\left\{\,\arg [t(p)]-\arg [t(p')] \,\right\}}{p-p'},
\ee
and this expression may be bounded by a term of order $\gd$.
Finally, the contributions of
$e^{-\I p_oy_o}\,\left[\,e^{-\I (p-p_o)\,y_o}-1\,\right]$ in this term
are of order $\gep(1+\gep)$ from eq.~(\ref{eq:Bornetnbis}) 
and the assumption on $\vert y_o\vert$ (since $\abs{(p-p_o)\,y_o}\ \leq \gep$).
\hfill{$\Box$}

\end{document}